\PassOptionsToPackage{table}{xcolor}
\documentclass[lettersize,journal]{IEEEtran}

\usepackage{booktabs}
\usepackage[bookmarksnumbered,unicode]{hyperref}

\usepackage{listings}
\usepackage{url}
\usepackage{color}
\usepackage{amsmath}
\usepackage{bm}
\usepackage{mathtools}
\usepackage{flushend}
\usepackage{multirow}
\usepackage{xcolor}
\usepackage{subfig}
\usepackage{mwe}
\usepackage[nounderscore]{syntax}
\usepackage{xspace}

\usepackage[english]{babel}
\usepackage{amsthm}

\theoremstyle{definition}
\newtheorem{definition}{Definition}[section]

\newcommand{\code}[1]{\texttt{#1}}

\newcommand{\thename}{{SmartFL}\xspace}
\newcommand{\thenamedetail}{{\textbf{S}e\textbf{M}antics b\textbf{A}sed p\textbf{R}obabilis\textbf{T}ic \textbf{F}ault \textbf{L}ocalization}}
\newcommand{\repo}{\textbf{\url{https://github.com/toledosakasa/SMARTFL}}}

\newcommand{\figref}[1]{Figure~\ref{#1}}
\newcommand{\tabref}[1]{Table~\ref{#1}}

\newcommand{\secref}[1]{Section~\ref{#1}}

\newcommand{\defref}[1]{Definition~\ref{#1}}
\newcommand{\smalltitle}[1]{{\smallskip \noindent \bf  {#1}.\ }}

\usepackage{listings}

\usepackage{hyperref}

\definecolor{boxgray1}{rgb}{0.95,0.95,0.95}
\usepackage[most]{tcolorbox}
\newtcbtheorem{Summary}{\bfseries Summary}{enhanced,drop shadow={black!50!white},
  coltitle=black,
  top=0.1in,
  attach boxed title to top left=
  {xshift=1.5em,yshift=-\tcboxedtitleheight/2},
  boxed title style={size=small,colback=boxgray1}
}{summary}

\definecolor{codegreen}{rgb}{0,0.6,0}
\definecolor{codepurple}{rgb}{0.58,0,0.82}
\definecolor{backcolour}{rgb}{0.95,0.95,0.92}
\definecolor{tablegray}{rgb}{0.88,0.88,0.88}
\newcolumntype{a}{>{\columncolor{tablegray}}l}
\lstdefinestyle{mystyle}{
    commentstyle=\color{codegreen},
    keywordstyle=\color{purple},
    identifierstyle=\color{blue},
    stringstyle=\color{codepurple},
    breakatwhitespace=false,         
    breaklines=true,                 
    numbers=left,                    
    numbersep=5pt,                  
    showspaces=false,                
    showstringspaces=false,
    showtabs=false,                  
    tabsize=4
}
\usepackage{amsmath,amsfonts}
\usepackage{algorithmic}
\usepackage{algorithm}
\usepackage{array}
\usepackage{textcomp}
\usepackage{stfloats}
\usepackage{url}
\usepackage{verbatim}
\usepackage{graphicx}
\usepackage{cite}
\begin{document}

\title{SmartFL: Semantics Based Probabilistic Fault Localization}


\author{Yiqian~Wu,
        Yujie~Liu,
        Yi~Yin,
        Muhan~Zeng,
        Zhentao~Ye,
        Xin~Zhang,
        Yingfei~Xiong,
        and Lu~Zhang

\IEEEcompsocitemizethanks{
\IEEEcompsocthanksitem Yiqian~Wu, Yujie~Liu, Yi~Yin, Muhan~Zeng, Zhentao~Ye, Xin~Zhang, Yingfei~Xiong, and Lu~Zhang are with Key Laboratory of High Confidence Software Technologies (Peking University), Ministry of Education; School of Computer Science, Peking University, Beijing, China.
\protect\\ Email: wuyiqian@pku.edu.cn, douya@stu.pku.edu.cn, yiyin2024@163.com, mhzeng@pku.edu.cn, ztye@pku.edu.cn, xin@pku.edu.cn, xiongyf@pku.edu.cn, zhanglucs@pku.edu.cn.
\IEEEcompsocthanksitem Xin Zhang is the corresponding author.
}

}



\maketitle

\begin{abstract}
Testing-based fault localization has been a research focus in software engineering in the past decades. It localizes faulty program elements based on a set of passing and failing test executions. Since whether a fault could be triggered and detected by a test is related to program semantics, it is crucial to model program semantics in fault localization approaches. Existing approaches either consider the full semantics of the program (e.g., mutation-based fault localization and angelic debugging), leading to scalability issues, or ignore the semantics of the program (e.g., spectrum-based fault localization), leading to imprecise localization results.
Our key idea is: by modeling only the correctness of program values but not their full semantics, a balance could be reached between effectiveness and scalability.
To realize this idea, we introduce a probabilistic model by efficient approximation of program semantics and several techniques to address scalability challenges.
Our approach, \thename(\thenamedetail), is evaluated on a real-world dataset, Defects4J 2.0. 
The top-1 statement-level accuracy of our approach is {14\%}, which improves 130\% over the best SBFL and MBFL methods. The average time cost is {205} seconds per fault, which is half of SBFL methods. After combining our approach with existing approaches using the CombineFL framework, 
the performance of the combined approach is significantly boosted by an average of 10\% on top-1, top-3, and top-5 accuracy compared to state-of-the-art combination methods.
\end{abstract}


\section{Introduction \label{sec:intro}}
In the last two decades, testing-based fault localization, or fault localization in short, has been a research focus in software engineering~\cite{CombineFL,sbfl_abreu2007accuracy,sbfl_wong2013dstar,mbfl_papadakis2015metallaxis,jiang2019combining}. Given a program and a set of tests with at least one failing test, a fault localization approach computes the suspiciousness score of each program element to determine which one is the most suspicious to be faulty. Here the program elements can be statements, methods, files, or any needed granularity.

Among the large body of fault localization research, a central focus is coverage-based fault localization.
Coverage-based fault localization infers the suspiciousness scores of program elements based on the coverage information, and the basic idea is 
that the fault location that causes a failing test should appear in the locations covered by the test and
an element covered more by failing tests rather than passing tests is more likely to be faulty. For example, spectrum-based fault localization (SBFL)~\cite{jones2002visualization}, one of the most well-known fault localization families, calculates the suspiciousness score of a program element based on the number of passing tests and the number of failing tests covering the element.

However, whether a buggy program element causes the failure of a test is determined by four conditions~\cite{voas1992pie,DBLP:conf/icst/SchulerZ11,DBLP:conf/sigsoft/ZhangM15}: (1) whether the test covers the buggy program element, (2) whether the execution of the buggy program element results in an error in the program state, (3) whether the error in the program state is propagated to the output, and (4) whether the error in the output is captured by an assertion or not. Coverage-based fault localization ignores the semantics of the target program and thus only considers the first condition. A test may cover a buggy program element but still pass because the latter three conditions are not satisfied, leading to inaccuracies in coverage-based fault localization.


To overcome this problem, different approaches have been proposed to take the latter three conditions also into consideration. For example, mutation-based fault localization (MBFL)~\cite{mbfl_moon2014ask,mbfl_papadakis2015metallaxis} generates many mutations on each element and watches whether the program output or the test result (i.e., the pass/fail status) changes. 
If a change in a statement is more likely to change the program output or the test result in the failing tests, and less likely in the passing tests, the statement is likely to be faulty.
Angelic debugging~\cite{chandra2011angelic} and semantic fault localization~\cite{DBLP:conf/tacas/ChristakisHMSW19} use symbolic analysis to determine whether the result of an expression can be modified to reverse the results of failing tests while maintaining the results of the passing tests, and such an expression is considered more likely to be faulty. However, these approaches take the full program semantics into consideration, and thus the analysis is inevitably heavy. As an existing study~\cite{CombineFL} reveals, mutation-based fault localization often requires hours to localize a single fault. As far as we are aware, there is so far no successful application of angelic debugging or semantic fault localization to large programs.

In this paper, we propose a novel approach to fault localization, \thename, that considers the four factors via efficient probabilistic modeling of the program semantics. Our approach considers a sample space of all possible faults and analyzes which program element is more likely to be faulty based on current test results. 
Our core insight is that the probability of a fault in the current program element leading to the current test results can be efficiently estimated by analyzing the following: 
\begin{itemize}
    \item the probability of each statement in the traces of test executions to introduce an error into the system state;
    \item the probability of each statement to propagate an error.
\end{itemize}

In this way, we do not need to consider the full semantics and can abstract each value into two possibilities: faulty or not. Consequently, the analysis is significantly simplified and can be efficiently approached. Along with this insight, we build a probabilistic model based on static and dynamic dependencies from the source code and test execution traces and calculate the posterior probabilities of whether a statement is faulty based on the test results.

However, realizing this idea still has scalability challenges. 
{Test executions can be long, and a model based on full test execution traces could be too large to build and to conduct inference.}
To overcome this challenge, there are solutions in two directions. The first solution is to reduce the size of the probabilistic model by discarding unimportant content. Specifically, this process consists of three parts, including reducing redundant methods, reducing redundant loops, and reducing redundant tests. The second solution is to propose a more efficient probabilistic inference algorithm. We design an optimized version of the probability inference algorithm~\cite{kschischang2001factor,DBLP:conf/aaai/Pearl82} specialized for{ inferring the posterior probabilities in} our model.

We have evaluated our approach on the widely-used Defects4J benchmark~\cite{just2014defects4j}. We experimented on all the 835 bugs
from Defects4J 2.0.
The results show that our approach significantly outperforms both SBFL and MBFL methods, in terms of both efficiency (205s per fault avg.) and effectiveness (14\% Top-1 statement accuracy). 
Our approach is also complementary to existing approaches: while combining our approach with existing approaches using the CombineFL framework~\cite{CombineFL}, the performance of the combined approach is
boosted by an average of 10\% on top 1, 3, and 5 accuracy
compared to other state-of-the-art approaches.

In summary, this paper makes the following main contributions.
\begin{itemize}
    \item A fault localization approach by efficient {approximation} of program semantics.
    \item Novel techniques {to reduce the size of the model and to efficiently infer posterior probabilities for addressing the scalability challenge.}
    \item An evaluation on the Defects4J dataset to show the effectiveness and the efficiency of our approach.
\end{itemize}

This paper is a significantly extended version of a previous conference paper ~\cite{DBLP:conf/icse/ZengWY0Z022}. 
\begin{itemize}
    \item First, this submission introduces new techniques to enhance the scalability of the SmartFL approach, including (1) adaptive folding, a technique for reducing redundant methods, (2) virtual call edge, a technique for capturing dependencies involving untraced methods, and (3) an optimized inference algorithm specialized for our models.
    \item Second, the implementation of SmartFL is improved for compatibility, stability, and efficiency, by designing a new version of tracing, better simulating the execution process of Java Virtual Machine, and supporting more Java features like run-time exceptions.
    \item Third, the evaluation has been extended to all the 835 bugs from Defects4J 2.0 and 
    the baselines are expanded. The
    results show that the new version of SmartFL is significantly improved compared to the old version in both effectiveness and efficiency. 
    \item Fourth, this submission has been largely rewritten to improve the presentation. In particular, we have switched from an informal approach description based on factor graphs to a formal description based on Bayesian networks for enhancing readability.
\end{itemize}

The rest of the paper is organized as follows. \secref{sec:overview} motivates our approach with
examples. \secref{sec:background} presents basic mathematical background about Bayesian networks.
\secref{sec:approach} describes our approach in detail.
\secref{sec:implementation} describes how to instantiate our approach for Java programs.
\secref{sec:evaluation} shows the experiment results and answers the research questions. 
\secref{sec:discussion} discusses some existing problems of our approach.
\secref{sec:related} discusses related research. 
\secref{sec:conclusion} concludes the paper.

\section{Overview \label{sec:overview}}


\begin{figure}
\footnotesize
\centering
\lstset{language=Java,basicstyle=\tt\footnotesize, linewidth=0.5\textwidth, xleftmargin=0.5cm}


\begin{lstlisting}[frame=single]
public class CondTest {
	public static bool foo(int a) {
		if (a <= 2) { // buggy, should be a < 2
			a = a + 1;
		}
		return a <= 2;
	}

	@Test
	void pass() {
		assertTrue(foo(1));
	}

	@Test
	void fail() {
		assertTrue(foo(2));
	}
}
\end{lstlisting}
\caption{A Motivating Example for Condition Modeling.}
\label{fig:example}
\end{figure}
In this section, we motivate our approach using an example.

\smalltitle{Motivating Example} 
\figref{fig:example} shows a simple program for illustration purposes. The buggy condition \code{a <= 2} at line 3 replaces the correct condition \code{a < 2}. There are two test cases to find the fault. For test \code{pass}, the fault does not influence the evaluation of the condition so the result is correct. However, in test \code{fail}, the fault misleads the test to the wrong branch and gets a wrong result. Here we assume a desirable approach should rank line 3 at the top in statement-level fault localization.

\smalltitle{Coverage-based Approaches}
We first demonstrate why coverage-based approaches such as SBFL fail to discover this bug. Coverage-based approaches utilize code coverage information to calculate suspiciousness scores. In SBFL approaches, the suspiciousness scores of an element $e$ are calculated from four numbers: the number of passing tests covering $e$, the number of failing tests covering $e$, the total number of passing tests, and the total number of failing tests. However, in the above case, the coverage of passing tests and failing tests are completely identical, resulting in equal suspiciousness scores for every statement regardless of specific SBFL formulas.

As analyzed in the introduction, SBFL formulas cannot distinguish the suspiciousness degrees of different statements because coverage is only one out of the four conditions that lead to test failure. In test \code{pass}, though the faulty expression is covered, the affected run-time state is still correct, and thus calculating suspiciousness with only coverage cannot distinguish each statement.

\smalltitle{Other Existing Approaches}
To address the above challenge, many existing approaches try to analyze also the latter three conditions, i.e., whether the execution of a statement produces a faulty state, whether the faulty state is propagated to the output, and whether the test captures the fault in the state. However, to analyze the three conditions precisely, we need to consider the full semantics of the program, which is difficult to achieve efficiently. Here we analyze two families of approaches.

A typical family is MBFL. MBFL mutates each statement to generate multiple mutants and checks whether the output of each test execution~\cite{mbfl_papadakis2015metallaxis} or the test result (i.e., the pass/fail status)~\cite{mbfl_moon2014ask} changes. In this case, mutating the statement at line 4 or the statement at line 6 has a high probability to fail test \code{pass}, while mutating the statement at line 3 has a much smaller probability to fail test \code{pass}. 
In this way, we know that the statement at line 3 has a weak correlation to the test result of \code{pass} and is more likely to be faulty. However, to obtain statistically significant information, we need to generate a number of mutants for each statement, and all tests need to be executed on each mutant, which takes a significant amount of time. In an existing empirical study~\cite{CombineFL}, mutation-based fault localization requires hours to localize a single fault.

Another representative family is angelic debugging~\cite{chandra2011angelic} and semantic fault localization~\cite{DBLP:conf/tacas/ChristakisHMSW19}. These approaches analyze, for each expression, whether its result can be modified to reverse the results of failing tests while maintaining the results of the passing tests. In this example, changing the result of expression \code{a+1} at line 4 or the result of expression \code{a} at line 6 to any value different from 2 would fail test \code{pass}, and thus the two expressions are not considered to be buggy. However, such an analysis requires symbolic reasoning, which is known to be heavy and has limited scalability. So far there is no successful application of angelic debugging or semantic fault localization to large programs within our knowledge.

\smalltitle{Our Approach}
Different from the above approaches, our approach takes a probabilistic view on fault localization. 
Let us consider a sample space of all possible faults that the current program could potentially contain. Given the current test results as an observation, our approach estimates the posterior probability of each program element being faulty.

{To precisely calculate the posterior probabilities, we need a distribution of the faults and to model the full semantics of the program. Specifically, first we need to find the prior probability distribution of the fault localization problem space, that is, the probability distribution of the potential correct versions of the program corresponding to the faulty program. Second, we need to analyze the result of each test on each correct version of the program and use the observed test results to calculate the posterior probability. 
Both tasks are difficult. It is difficult to know the prior distribution of the potential correct versions and it is difficult to analyze many program versions based on their full semantics, which requires probabilistic modeling of the relationships between specific program states. 
These difficulties make such probabilistic modeling impossible to achieve, and even if it is possible, it will face serious effectiveness and efficiency problems.
The lack of reasonable prior probability leads to poor performance. Even if the prior problem is overcome, the model will have serious efficiency problems in inference due to modeling the complete semantics.
}

To overcome the above difficulties, our approach takes a core assumption: a faulty evaluation results in an evenly distributed random result. Specifically, an evaluation of an expression is faulty only if one of the three conditions is satisfied.
  \begin{itemize}
  \item The expression itself is faulty.
  \item Any input value of the expression is faulty.
  \item The expression was executed by mistake; in a correct execution it should not be executed.
  \end{itemize}
When any of the above three conditions hold, we assume that the expression would produce an evenly distributed random output. This assumption allows us to use Bernoulli random variables to model whether values in execution are correct or not; we do not have to find the prior distribution of the potential correct versions, nor do we need to consider the full semantics of the potential correct versions in the analysis. Through this modeling, we not only scale the efficiency of probabilistic inference but also simplify the prior parameters.


Concretely, we introduce a set of Bernoulli random variables to represent whether a statement is correct, denoted by $S_i$, where $i$ is the line number of the statement. It is natural to assume these random variables are independent of each other. We also introduce another set of Bernoulli random variables to represent whether the output value of an expression execution is correct. In this example, we use $V_{p, i}$ ($V_{f, i}$) to denote the value produced by the expression execution at line $i$ in test \code{pass} (\code{fail}). In particular, $V_{p, 2}$ and $V_{f, 2}$ denotes the correctness of the test inputs and $V_{p, 6}$ and $V_{f, 6}$ denotes the correctness of the test outputs. 



{Now let us analyze the relations between the random variables. First, since}
 the input values of the tests are correct, we have the following equations.
\[
    P(V_{p, 2} = 1) = 1, P(V_{f, 2} = 1) = 1
\]

Please note that since the Bernoulli random variables are binary, we also know $P(V_{p, 2} = 0) = P(V_{f, 2} = 0) = 0$. To ease the presentation, we will only present one of the two probabilities.


{
Then we consider the case of correct executions. Given a statement, if the statement itself is correct, the input values of the statement are correct, and the statement should be executed, then the result of the statement must be correct. As a result, we have the following equations:
}
\[
    \begin{array}{l}
        P(V_{t, 3} = 1 \mid S_3 = 1\wedge V_{t, 2} = 1) = 1 \\
        P(V_{t, 4} = 1 \mid S_4 = 1\wedge V_{t, 2} = 1\wedge V_{t, 3} = 1) = 1 \\
        P(V_{t, 6} = 1 \mid S_6 = 1\wedge V_{t, 4} = 1) = 1 \\
    \end{array}
\]
$\textrm{where }t\in\{p, f\}$.

{The first equation corresponds to the conditional expression in line 3. The correctness of the expression is represented by $S_3$. The expression reads the input parameter \code{a} in line 2, whose correctness is represented by $V_{t, 2}$. The statement always needs to be executed, so we do not need additional conditions. The second equation corresponds to the statement in line 4. Similarly, the correctness of the statement is represented by $S_4$ and the statement reads the input parameter \code{a} whose correctness is represented by $V_{t, 2}$. Yet whether the statement should be executed depends on the result of the conditional expression in line 3, and thus we need to include $V_{t, 3}$ in the condition part. The last equation is similar.}


Next, we consider the case where faulty values may be produced or propagated. 
{
The result of an expression could be incorrect if any type of a faulty execution occurs: (1) the expression is incorrect, (2) the input is incorrect, and (3) the expression is executed by mistake. However, a faulty execution does not necessarily result in an incorrect result. In general, the probability of a faulty execution leading to a faulty result is difficult to calculate as it depends on the type of fault and the previous system state. However, based on our assumption, 
all faulty evaluation results in an evenly distributed random output, and thus we ignore these dependencies and calculate the probability of the output being faulty based on the domain of the result.
If the size of the domain is $n$, the probability of a result being faulty is $(n-1)/n$. As a result, we have the following equations:
}

\[
    \begin{array}{l}
        P(V_{t, 3} = 0 \mid S_3 = 0\vee V_{t, 2} = 0) = 0.5 \\
        P(V_{t, 4} = 0 \mid S_4 = 0\vee V_{t, 2} = 0\vee V_{t, 3} = 0) = 0.99 \\
        P(V_{t, 6} = 0 \mid S_6 = 0\vee V_{t, 4} = 0) = 0.99 \\
    \end{array}
\]
$\textrm{where }t\in\{p, f\}$.

{
Our assumption also gives us an additional benefit: isolating the random variables for different expressions. In the general case of using the full semantics, even if we know the input for line 6 is faulty, whether the output of line 6 is faulty still depends on other factors. For example, if line 3 is correct and line 4 is faulty, and the correct version is \code{a=5}, then in test \code{fail}, while the result of line 4 is faulty (should be 5 but is actually 3), the result of line 6 is correct. On the other hand, if the faulty line is line 3 as in our example but not line 4, the results of line 4 and line 6 are both incorrect in test \code{fail}. In other words, $V_{t, 6}$ depends on $S_3$ and $S_4$ even when $V_{t, 4}=0$. Such indirect dependencies will make the model strongly coupled and the calculation of posterior probabilities difficult. On the other hand, with our assumption, the correctness of an expression result only depends on the random variables that directly affect the expression. 
}

{Furthermore, for probability calculation, we need to know the prior probabilities of statements being faulty. Since we do not have any prior knowledge about their faulty probability, we simply assume the same prior probability for all statements.
\[
    P(S_3 = 1) = 0.5, P(S_4 = 1) = 0.5, P(S_6 = 1) = 0.5
\]}

\begin{figure}[t]
  \centering
  \includegraphics[width=0.5\textwidth]{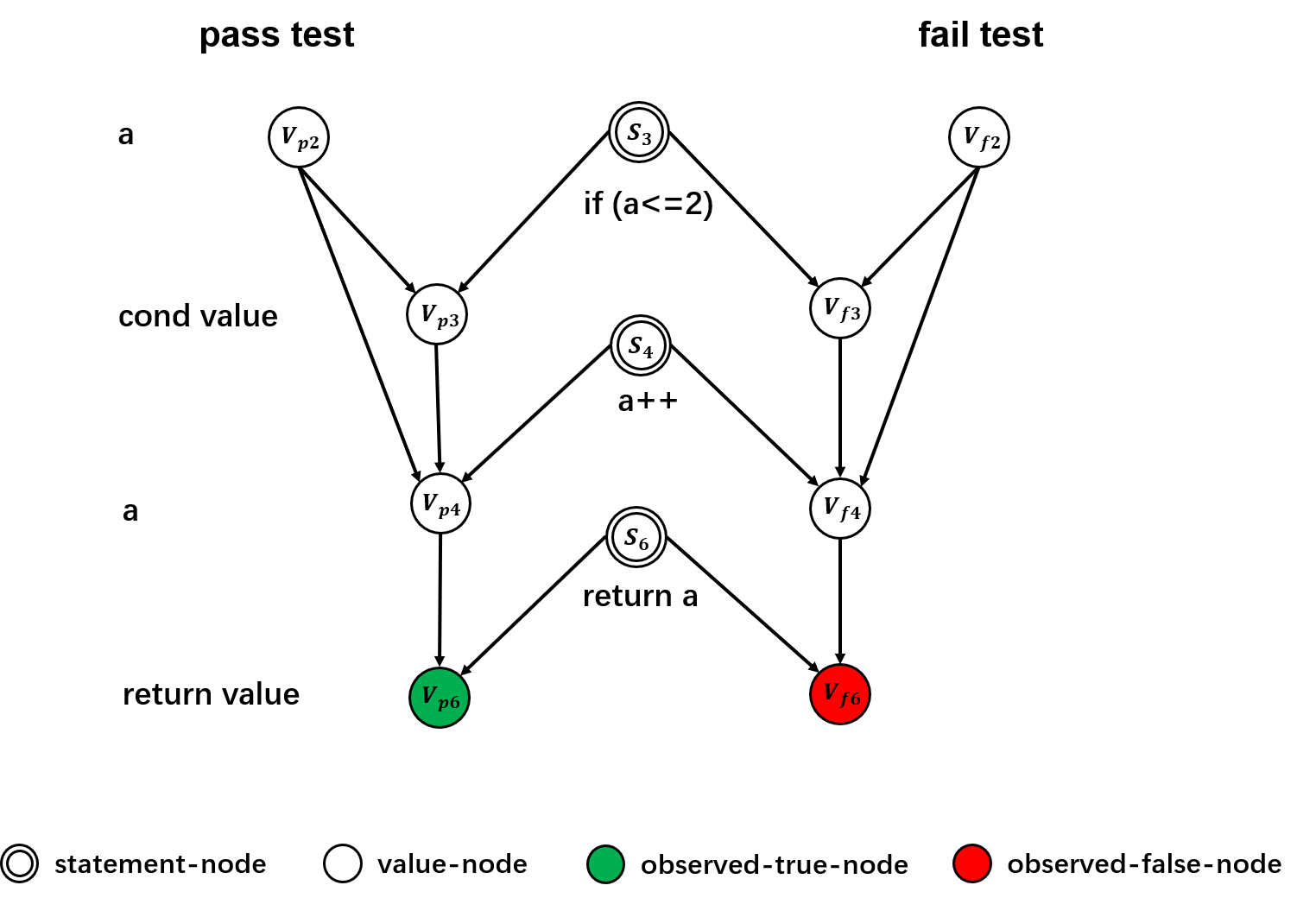}
  \caption{Generated Bayesian Network for Figure~\ref{fig:example}.}
  \label{fig:example_DAG}
\end{figure}

{
Based on the above analysis, we can model the relations between the random variables as a Bayesian network, as shown in \figref{fig:example_DAG}, where the nodes are random variables, and edges are conditional dependencies. The details of Bayesian networks will be introduced in \secref{sec:background}.
Based on the definition of the Bayesian network, we can calculate the joint probabilities from the conditional probabilities represented by the edges. Let $Pr = P(V_{t, 2}, V_{t, 3}, V_{t, 4}, V_{t, 6}, S_3, S_4, S_6)$ be the joint probabilistic distribution, we can represent it as follows:
\[
    \begin{aligned}
    Pr = & P(V_{t, 2})P(S_3)P(V_{t, 3}\mid S_3, V_{t, 2}) \\
        & P(S_4)P(V_{t, 4}\mid S_4, V_{t, 2}, V_{t, 3}) \\
        & P(S_6)P(V_{t, 6}\mid S_6, V_{t, 4})\}
    \end{aligned}
\]
$\textrm{where }t\in\{p, f\}$.
The above formula allows us to infer probabilities over this model. There also exist approximating algorithms to efficiently infer probabilities on large models~\cite{DBLP:conf/aaai/Pearl82}.
}

{Finally, based on the test results, we have the following observations.}
\[
    V_{p, 6} = 1,  V_{f, 6} = 0
\]
{Based on this observation, we can calculate the posterior probability of each statement being faulty. Using the loopy belief propagation algorithm~\cite{DBLP:conf/aaai/Pearl82} on this model, we can infer the following probabilities.
\[
\begin{aligned}
   P(S_3=0\mid V_{p, 6} = 1,  V_{f, 6} = 0) \approx 0.707\\
   P(S_4=0\mid V_{p, 6} = 1,  V_{f, 6} = 0) \approx 0.270\\
   P(S_6=0\mid V_{p, 6} = 1,  V_{f, 6} = 0) \approx 0.223 \\
\end{aligned}
\]
Therefore, $S_3$ has the highest probability to be faulty, i.e., successfully localizing the fault.}

\smalltitle{Challenges and Solutions} 
While the basic idea is straightforward, realizing this idea still faces a scalability challenge: the Bayesian network built may be too large to be solved efficiently.  
We address this challenge from four aspects. 
More details can be found in ~\secref{sec:performance}.
\begin{itemize}
\item First, we find a project may contain many tests, and modeling all of them may lead to a very large model, while many tests are unrelated to the current fault. To address this issue, we introduce a two-phase instrumentation and use a coarse-grained instrumentation to filter out tests unrelated to the failing ones.
\item Second, we find a test execution trace may be extremely long due to the existence of loops (or recursions), while such long loops provide repeated information, and modeling such a trace alone leads to a very large model. To address this issue, we introduce a loop compression algorithm to select typical iterations such that all control/data dependencies between statements and variables within any iteration are covered by at least one selected iteration. In this way, we model the main effects of the loop execution with a small number of iterations.
\item Third, we find some methods are not critical to the faults, such as those that are only covered by passing tests
(If a method is only covered by passing tests and not by failing tests, then the method does not contain any statement that triggers a fault in a failing test). 
To address this issue, we introduce a technique called ``partial tracing'' to ignore the specific details in some methods. To avoid losing the dependencies involved in these untraced methods, we design a technique to model these dependencies.
\item Fourth, we find that in the general probabilistic inference algorithm (loopy belief propagation), the time cost of a calculation for a term increases exponentially with its degree (the number of incident nodes). To address this issue, we design a specialized version of the inference algorithm for our probabilistic model. Because the conditional probabilities in our probability model all meet a unified format and can be represented in a compressed way, we do not need to traverse each table item and can manually derive the calculation formula by extracting common factors, which simplifies the exponential calculation process to a constant level.
\end{itemize}

\section{Background \label{sec:background}}

Before introducing our approach, we describe background information about Bayesian networks, on which our probability model is based.

A Bayesian network is a probabilistic graphical model representing a set of random variables and their conditional dependencies via a directed acyclic graph (DAG).
Formally, a Bayesian network is a DAG $G = (V, E)$ together with a random variable $x_i$ for each node $i\in V$ and one conditional probability distribution $p(x_i | \textbf{x}_{A_i})$ per node, specifying the probability distribution of $x_i$ conditioned on its preceding variables $\textbf{x}_{A_i}$\footnote{We use bold fonts to represent a set.} in the graph. 
Thus, a Bayesian network defines a joint probability distribution, by multiplying all the conditional probabilities.
A Bayesian network also allows specifying posterior distributions by incorporating evidences.
The form of evidences we consider in the paper is the one that fixes a random variable to a value.
In other words, if the original Bayesian network defines a joint distribution $p(\textbf{x})$, and the evidence is $x_i = v$, the the posterior distribution is $p(\textbf{x} \mid x_i = v)$.

The advantage of defining a probability distribution using a Bayesian network is that it can decompose the joint probability distribution into a product of factors, which allows inference algorithms to compute the marginal distribution efficiently.
In particular, loopy belief propagation~\cite{DBLP:conf/aaai/Pearl82} is an efficient approximate algorithm to infer the marginal distributions, which we apply in our implementation.

To illustrate the procedure of loopy belief propagation, we denote $F$ to be the set of conditional probability distributions.
Each probability distribution in $F$ is called a factor. 
Intuitively, the random variables and the factors also form an undirected graph in which they are nodes.
For a conditional probability distribution $p(x | x_{A})$, all random variables in $x_{A} \cup \{x\}$ are incident to this factor in the graph.
In other words, 
there is an edge between a factor and a random variable if and only if the random variable is involved in the corresponding conditional probability distribution.
The joint probability distribution can be expressed as:
\[
    p(\mathbf{x}) = \prod_{a \in F} f_{a}(\mathbf{x}_{a})
\]
where $a$ denotes a factor, $f_{a}$ denotes the conditional probability function of $a$ and $\mathbf{x}_{a}$ denotes the vector of incident random variables of $a$.
The graph is known as a \emph{factor graph}~\cite{DBLP:journals/tit/KschischangFL01}.

Loopy belief propagation is a multi-round iterative message-passing algorithm.
It iteratively updates messages that go between a factor and a random variable that are incident, which are in turn used to calculate the marginal probabilities in the end.
These messages are real valued functions that map the value of a given random variable to a real value.
We use $\mu_{v \rightarrow a}^n$  to denote the message that goes from a random variable $v$ to a factor $a$ in the $n$th iteration and $\mu_{a \rightarrow v}^n$ to denote the message that goes in the opposite direction.

Each iteration is carried out alternately in two steps. In the first step, each random variable sends a message to all its incident factors.
This message is the product of the messages received by the random variable from other incident factors (except the factor receiving the message). A message $\mu_{v \rightarrow a}^n(x_v)$ from a random variable $v$ to a factor $a$ is defined by:
\begin{equation}
    \mu_{v \rightarrow a}^n(x_v) = \prod_{a^{*} \in I(v)\backslash \{a\}} \mu_{a^{*} \rightarrow v}^{n-1}(x_v) \label{eq:vf}
\end{equation}
where 
$I(v)$ denotes the set of incident factors of $v$.

In the second step, each factor sends a message to all its incident random variables. This message is computed using the messages received from other incident random variables (except the node receiving the message being computed) and
the conditional probability function of the factor.
A message $\mu_{a \rightarrow v}^n(x_v)$ from a factor $a$ to a random variable $v$ is defined by:
\begin{equation}
    \mu_{a \rightarrow v}^n(x_v) =  \sum_{\mathbf{x}_{a} \backslash x_v} 
    \left(f_{a}(\mathbf{x}_{a}) \prod_{v^{*} \in I(a)\backslash \{v\}} \mu_{v^{*} \rightarrow a}^{n}(x_{v^{*}})\right) \label{eq:fv}
\end{equation}
where $I(a)$ denotes the set of incident variables of $a$, $\mathbf{x}_{a}$ denotes the values of the set of random variables that are incident to $a$ (including $x_v$),  and the summation $\sum_{\mathbf{x}_{a} \backslash x_v}$ means to sum over all possible values in $\mathbf{x}_{a}$ while the value of $v$ is fixed to $x_v$.

To improve convergence, loop belief propagation implementations typically apply normalization on the calculated messages in each iteration:
\[ \mu_{v \rightarrow a}^n(x_v) = \mu_{v \rightarrow a}^n(x_v) / \sum_{c \in domain(v)} \mu_{v \rightarrow a}^n(c),\]
\[
\mu_{a \rightarrow v}^n(x_v) = \mu_{a \rightarrow v}^n(x_v) / \sum_{c \in domain(v)}  \mu_{a \rightarrow v}^n(c).\]

The algorithm iterates until the messages converge or the iteration exceeds a certain time limit. Suppose the algorithm runs for $N$ iterations, the marginal probability of each random variable is calculated using the product of all final messages from incident factors:
\[
    p(x_v) = \frac{\prod_{a \in I(v)} \mu_{a \rightarrow v}^N(x_v)}
    {\sum_{x_v} \prod_{a \in I(v)} \mu_{a \rightarrow v}^N(x_v)}
\]
where the denominator is a normalization constant.

\section{Our Approach}
\label{sec:approach}

\begin{figure}[t]
  \centering
  \includegraphics[width=0.5\textwidth]{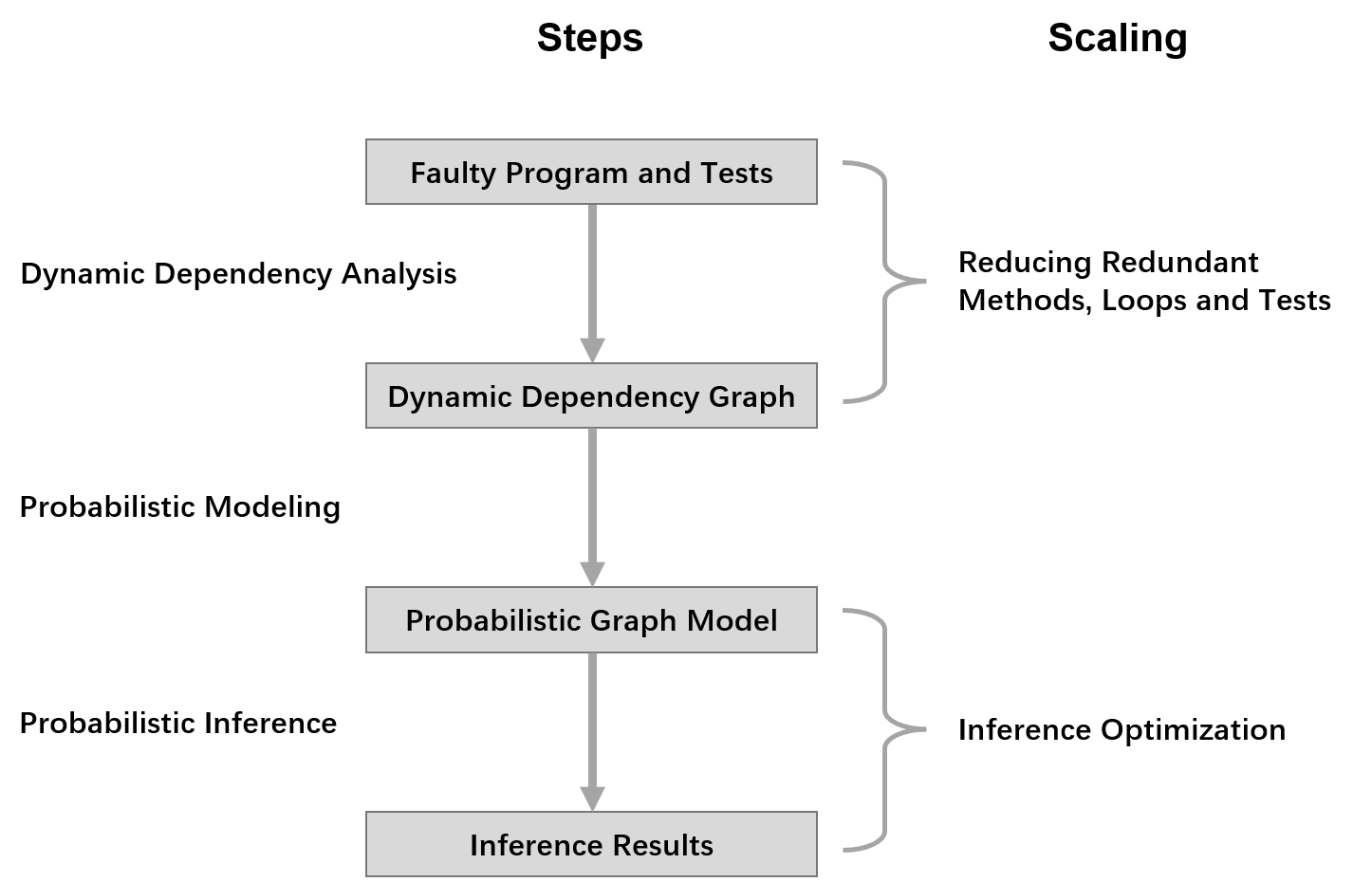}
  \caption{Approach Workflow.}
  \label{fig:framework}
\end{figure}

Figure~\ref{fig:framework} outlines the workflow of our approach.
The input is a program and a set of test cases.
First, our approach converts the input into a dynamic dependency graph which describes the data dependencies and control dependencies between statements and values in the test cases by instrumenting the test cases and performing a static analysis.
Then our approach converts the dynamic dependency graph into a Bayesian network.
The network defines a joint distribution of whether each statement and value in the test executions is faulty.
Finally, by conditioning on the results of the test cases, our approach performs marginal inference on the Bayesian network and ranks the statements by the marginal probabilities of them being incorrect.
To scale our approach, we apply various optimizations to reduce the dynamic dependency graph size and improve the inference algorithm efficiency.

In the rest of the section, we go into each step of our approach in detail with an emphasis on how to convert the dependency graph into a Bayesian network, and describe our optimization techniques in the end.

\subsection{Dynamic Dependency Analysis}
\label{sec:DDG}
The first step of our approach is to build a dynamic dependency graph that captures data dependencies and control dependencies in the test cases.
Abstractly, the program execution on a test case can be viewed as a sequence of statements along with input values of the test case and values that are produced by the statements.
A statement is a static notation and can appear multiple times in an execution.
We refer to one statement's appearance as a statement instance.
On the other hand, a value is a dynamic notation and can be only produced by one statement instance in a test case.
Each statement instance can be viewed as a function that maps a set of values to a value.
We say there is a data dependence from a statement to a value if one of the statement instances produces the value.
There is also a data dependence from a value to another value if the latter is produced using the former.
We say there is a control dependence from a value to a statement instance if the value is used in an instance of a control statement (e.g., a branching statement) that the statement is control dependent on.
How to obtain such dependencies depends on the programming language.
In Section~\ref{sec:implementation}, we describe it for Java.
Briefly, our approach obtains the data dependency by instrumenting the test executions and performing a dynamic analysis~\cite{DBLP:conf/isca/AustinS92};
our approach obtains the control dependency by performing a static analysis~\cite{cooper2001simple} that calculates the dominance relations~\cite{dragonbook} between statements.

We build a dynamic dependence graph for all test cases in the input using the aforementioned data dependency and control dependency:
\begin{definition}[Dynamic Dependency Graph]
\label{def:ddg}
Given a set of test cases on a program and their data/control dependencies, a dynamic dependency graph is a directed acyclic graph $G = (N,E)$.
There are two kinds of nodes in $N$:
\begin{enumerate}
    \item For each statement $s$ in the program, there is a unique statement node $n_s \in N$.
    \item For each value $v$ in the test cases, there is a unique value node $n_v \in N$.
\end{enumerate}
There are three kinds of edges in $E$:
\begin{enumerate}
    \item For each data dependence from an instance of statement $s$ to a value $v$, there is an edge from the corresponding statement node $n_s$ to the corresponding value node (i.e., $\langle n_s, n_v\rangle \in E$).
    \item For each data dependence from a value $v$ to a value $v'$, there is an edge between the corresponding value nodes (i.e., $\langle n_v, n_{v'}\rangle \in E$).
    \item For each control dependence from a value $v$ to a value $v'$, there is an edge between the corresponding value nodes (i.e., $\langle n_v, n_{v'}\rangle \in E$).
\end{enumerate}
\end{definition}

\paragraph{Example.} Consider the Bayesian network in Figure~\ref{fig:example_DAG}.
Ignoring the conditional probability distributions, its graph structure is the dynamic dependency graph for the example program in Figure~\ref{fig:example} with two test cases.
Nodes $S_3$, $S_4$, and $S_6$ are three statement nodes and other nodes are value nodes.
Note the statement nodes are static and shared between the two test cases.
On the other hand, the value nodes are dynamic and specific to each test case.
The edges $\langle S_4, V_{p_4}\rangle$, $\langle V_{p_3}, V_{p_4}\rangle$, and $\langle V_{p_2}, V_{p_4}\rangle$  correspond to the three kinds of edges respectively.

\subsection{Probabilistic Modeling}\label{sec:pgm}
After building the dynamic dependency graph, our approach transforms it into a Bayesian network.
The Bayesian network has the same structure as the original dependency graph and defines a joint distribution of whether each statement and each value can go wrong.
In addition, the Bayesian network incorporates the test results by encoding them as evidences.
We next describe how our approach constructs the Bayesian network in detail.

\paragraph{Graph Structure} For each statement node and each value node in the dynamic dependency graph, there is a node and a Bernoulli random variable representing whether the statement or value is faulty.
For two random random variables, there is an edge between them if and only if there is an edge between their corresponding nodes in the dynamic dependency graph.

\paragraph{Conditional Probabilities} For statement random variables, they have no parents in the Bayesian network.
We set a uniform prior probability 0.5 for all of them (i.e., $p(x_s) = 0.5$ for a statement variable $x_s$).
As for a value random variable, it has no parents if and only if it is an input value.
We assume input values are always correct, so we set their prior probabilities to 1 (i.e., $p(x_i) = 1$ for an input value variable $x_i$).
As for a value that is produced by a statement, \emph{the idea is that it can be erroneous if the statement is faulty or any value that is used in the statement is erroneous; otherwise, it must be correct}.
We model this idea using the conditional probability between the value variable and its parents, which is the key to our abstraction and modeling of program semantics.
For a value that is produced by a statement, let its random variables be $x_v$, and the random variables corresponding to its parents be $x_1,...,x_n$, then we have
\[
p(x_v = true \mid x_1\wedge ... \wedge x_n)  = 
\begin{dcases}
1,\quad x_1\wedge ... \wedge x_n = \textit{true} \\
p_0,\quad x_1\wedge ... \wedge x_n = \textit{false}
\end{dcases}.
\] 
Here $p_0$ is a hyper-parameter.
The idea is that even if the statement is faulty or any value it uses is erroneous, there is still a chance that the produced value is correct.
For example, let the correct statement be $a \leq 1$ where $a=0$, even if the statement becomes $a<1$ evaluating it still produces $true$.
Further, we observe that statements whose operators are ``\code{<}'', ``\code{>}'', ``\code{==}'', ``\code{!=}'' or ``\code{\%}'' are less sensitive to the correctness of the parents than other operators (e.g. ``+'', ``$\times$'').
For example, when $x$ contains an incorrect value, there is a higher chance for $x > 0$ to produce a correct value than $x+1$, as the prior has a much smaller value domain $\{True, False\}$.
As a result, we set $p_{0}$ differently for these two types of statements.
For statements with a wide range of the domain like ``\code{+}'', we set $p_{0} = 0.01$, which means that the result only has a very low chance of being correct. For statements with a boolean range like ``\code{<}'', we set $p_{0} = 0.5$, which means the result is equally likely to be correct or incorrect.  We refer to parameters 0.01, and 0.5 as very low, and moderate probabilities in \secref{sec:evaluation}. The impact of different parameter values is further discussed in RQ4 of \secref{sec:evaluation}.

\paragraph{Evidences} Finally, we encode the test results as evidences in the Bayesian network.
The test results are reflected by the boolean values used in the assertions (e.g., evaluation of ``\code{x>0}'' in ``\code{assert(x>0)}'').
Let $\textbf{x}_t$ be the set of random variables that correspond to the values that are used in passing assertions and $\textbf{x}_f$ be the set of random variables that are used in failing assertions.
Then $\bigwedge_{x\in \textbf{x}_t} x = true$ and $\bigwedge_{x\in \textbf{x}_f} x = false$ are added as evidences to the Bayesian network.
As a result, the final Bayesian network defines a joint distribution of whether each value or statement is correct given the test results:
\[P(\ \textbf{x}_S, \textbf{x}_V \mid \bigwedge_{x\in \textbf{x}_t} x = \textit{true} \wedge \bigwedge_{x\in \textbf{x}_f} x = \textit{false}\ ),\]
where $\textbf{x}_S$ is the set of random variables that correspond to statements, and $\textbf{x}_V$ is the set of random variables that correspond to values.


\subsection{Probabilistic Inference}
Once the Bayesian network is built, our approach computes the marginal probability of a statement being correct, which is used to rank which statements are more likely to be faulty.
For a statement $s$, let its corresponding random variable be $x_s$, then its marginal probability of being correct $P(x_s = true )$ can be calculated as:
\[\sum_{\textbf{x}_S\cup \textbf{x}_V \setminus \{x_S\}  }P(\ \textbf{x}_S, \textbf{x}_V \mid \bigwedge_{x\in \textbf{x}_t} x = \textit{true} \wedge \bigwedge_{x\in \textbf{x}_f} x = \textit{false}\ ).\]
We implement the inference using loopy belief propagation, as described in ~\secref{sec:background}.

\subsection{Scaling Our Approach}
\label{sec:performance}
The size of the Bayesian network
depends on the size of the dynamic traces used to build the graph.
As described in \secref{sec:overview}, instrumenting program runs can produce enormous execution traces, which can lead to gigantic probabilistic graphs that cannot even be stored in physical memory, let alone performing inference.
To address this challenge, we apply several techniques to exclude certain test cases and reduce the trace sizes. Also, we propose an optimized inference algorithm to speed up the probability inference.

\subsubsection{Reducing Redundant Tests}
\label{sec:selecting}
Our approach identifies and excludes test cases that are unlikely to contain useful information or greatly hinder the efficiency of our approach.

First, we observe that while all failing test cases carry useful information about faults, not all passing test cases are useful, some of which can be excluded.
The intuition is that a failing test must execute at least one faulty statement, while for a passing test, the more it overlaps with passing tests in terms of coverage, the more likely it executes faulty statements.
Since a faulty statement must have been executed by a failing test to trigger the fault, if a passing test has no overlap with any failing test in coverage, then it carries no information for fault localization as it does not cover any faulty statement that triggers a fault.
Based on this intuition, before generating dynamic dependency graphs, we first run lightweight instrumentation to get method-level coverage for each test case, and then rank the passing tests based on the number of methods that are both covered by the failing test and any passing test.
Finally, we only keep the top 50 passing tests for generating dynamic dependency graphs and exclude the rest for efficiency.


After the above test selection process, the resulting probabilistic graph can be still too large so we further exclude test cases that can blow up the graph size.
First, in the instrumentation phase, we set a hard limit on the size of the produced trace, and if the size of any trace exceeds the limit, we discard the corresponding test case.
Second, we set a hard limit on the size of the probabilistic graph under construction, and when the size exceeds the limit, we discard all the passing test cases that have not been considered yet.
Concretely, we consider all failing test cases when constructing the graph, following the intuition that failing test cases are more likely to carry useful information.
If the graph size has already exceeded the limit, we do not consider any passing test case.
Otherwise, we sort the passing test cases in an ascending order based on their trace sizes and incorporate them one by one into the graph until the graph size exceeds the limit. 

\subsubsection{Reducing Redundant Methods}
\label{sec:partial}
Besides excluding certain test cases, our approach applies techniques to compress traces produced by the test cases at different granularity.
Note in the aforementioned technique where our approach excludes test cases of large trace sizes, the trace sizes are calculated after compression.
The first technique is to compress redundant methods into atomic statements. 
A key challenge is how to maintain important dependencies after compression.
We next introduce which methods to compress and how to address the challenge.

Similar to the previous technique to reduce test cases, our approach compresses methods that are unlikely to contain a faulty statement and methods that would lead to long traces.
For the former kind of method, our insight is that faulty statements that are responsible for the failure must be in the application code (i.e., not in libraries), and be covered by at least one of the failing test cases, which means our approach needs to trace only the application methods covered by failing tests.
We refer to this technique as ``partial tracing''.

For the latter kind of method, our approach compresses large methods in large traces. 
As mentioned in the previous subsubsection, our approach excludes test cases whose sizes exceed a limit.
However, this is undesirable for failing test cases as they carry important information about faulty statements.
So instead of excluding such test cases, we propose a technique called ``adaptive folding'' to compress them.
Concretely, the technique sorts each method in a failing test case whose trace size exceeds the limit from large to small according to the number of statement instances it comprises in the trace and then compresses methods one by one until the trace size is below the limit.
While partial tracing avoids tracing certain methods at the instrumentation stage, adaptive folding is a post-processing technique after a trace is generated.

Now we describe how to maintain dependency integrity after compression, especially given our approach has avoided tracing certain methods in partial tracing. We face two main challenges: how to handle side effects and how to handle callbacks to uncompressed methods.
As the situation varies for different invocations to the same method, we discuss our solutions at method invocation level.
For method invocations not involving any of the two cases, they can be compressed into atomic statement instances which read their parameters and write to their return values.
When considering side effects, a method invocation may read or write any memory location that is reachable via global variables (static fields in Java programs) or parameters of reference type.
However, following this conservative assumption would lead to a lot of false dependencies which would degrade the performance of our approach severely.
To balance soundness and completeness, we assume any method invocation does not access global variables (which is often the case in practice for Java programs), and each invocation only access a parameter object's fields but not other memory locations that are transitively reachable via fields.
In other words, we abstract each object as a memory location and when its reference is used as a parameter of a method invocation, the invocation is considered to both read and write this object.

When an untraced method invokes a traced method, if we directly compress the untraced invocation, we would lose its data dependency with the traced invocation. 
To address this challenge, we introduce ``virtual call edges'' between the atomic statement instance that is produced by compressing the untraced invocation and the traced invocation.
Concretely, we assume the statement instance writes to the traced invocation's parameters and reads its return value.

\subsubsection{Reducing Redundant Loops}
\label{sec:compress}
The second technique to compress a trace is to reduce redundant loops, which we refer to as ``loop compression''. The core idea of this technique is to remove repeated loop iterations from the trace to get a compressed trace.
The key intuition is that the dependency information a long-running loop provides can be captured by a few iterations of it as the patterns of most iterations are repetitions of those of these few key iterations.
In particular, if the sequence of statement instances in one iteration is identical to that in another, the iteration is redundant.
This is because the subgraphs introduced by these two iterations in the dynamic dependency graph are isomorphic, and prove the same dependency information.
We identify and remove such redundant iterations by checking every pair of adjacent iterations in a loop, and removing the latter iteration if the two iterations comprise the same sequence of statement instances.
For example, a sequence like ``$\mathrm{(ab)^{100}ad(ab)^{100}}$'' is compressed into ``abadab''.
For nested loops, we first compress the inner loops and then compress the outer loops. 



\subsubsection{Inference Optimization}
\label{sec:optimization}
Besides compressing the probabilistic graphical model, we also propose a new optimized inference algorithm.
Concretely, by exploiting the structure of our conditional probability constraints, we are able to reduce a computation that is exponential in the number of variables involved in a conditional probability distribution into a linear one, which greatly improves the overall efficiency of the inference.

To infer the marginal probability of each program statement being incorrect, we implement loopy belief propagation, which is described in \secref{sec:background}. 
A key efficiency bottleneck in the algorithm is the computation of a message that is sent from a factor to a variable (see Equation~\ref{eq:fv}), which is exponential in the number of random variables that are incident to the factor.
Concretely, the algorithm needs to enumerate all possible assignments to the variables involved.
In our case, a factor is a conditional probability distribution and all random variables are Bernoulli, so if $d$ variables are involved in a conditional probability distribution, the complexity of computing a message would be $2^{d-1}$.

In order to overcome this problem, we design an optimized version of the inference algorithm by exploiting the structure of our conditional probability distributions. 
The core idea of this optimization is that in our scenario, for a given conditional probability distribution, many different assignments yield the same probability, so we can group their computation together and avoid explicitly enumerate all assignments. By utilizing this idea, we can simplify the computation of a message from exponential complexity into linear complexity, which we will introduce in detail below.

As described in Section~\ref{sec:pgm}, a conditional probability in our modeling is in the form of 
\[
p(x_v = true \mid x_1\wedge ... \wedge x_n)  = 
\begin{dcases}
1,\quad x_1\wedge ... \wedge x_n = \textit{true} \\
p_0,\quad x_1\wedge ... \wedge x_n = \textit{false}
\end{dcases},
\]
where $V=\{v_1, v_2, \dots, v_n\}$ are parent random variables of random variable $v$ in the Bayesian network, $x_v$ is the value of $v$, and $x_i$ is the value of $v_i$.
For convenience, we denote $y = x_v$  and $X = \{x_1, x_2, \dots, x_n\}$. We denote the factor that corresponds to the conditional probability as $a$ and then its function $f(y,X)$ can be expressed as:
\begin{align*} 
f(\textit{true},X) &= 1, &\forall x_i \in X: x_i = \textit{true} \\
f(\textit{false} ,X) &= 0, &\forall x_i \in X: x_i = \textit{true} \\
f(\textit{true},X) &= p_0, &\exists x_i \in X: x_i = \textit{false} \\
f(\textit{false} ,X) &= 1-p_0, &\exists x_i \in X: x_i = \textit{false}  \\
\end{align*}

The message from the factor $a$ to the random variable $v$ is:
\[
    \mu_{a \rightarrow v}(y) = \sum_{X} f(y,X) \prod_{v_i \in V} 
    \mu_{v_i \rightarrow a}(x_i)
\]
In addition, in the loopy belief propagation process, normalization is performed at each iteration, that is $\mu_{v \rightarrow a}(\textit{false}) + \mu_{v \rightarrow a}(\textit{true}) = 1$. As a result, we have:
\begin{align*} 
\mu_{a \rightarrow v}(\textit{true}) &=\sum_{X} f(\textit{true},X) \prod_{v_i \in V} \mu_{v_i \rightarrow a}(x_i) \\
 & = (1-p_0) \prod_{v_i \in V} \mu_{v_i \rightarrow a}(\textit{true}) + p_0\\
\end{align*}
\begin{align*} 
\mu_{a \rightarrow v}(\textit{false}) &=\sum_{X} f(\textit{false},X) \prod_{v_i \in V} \mu_{v_i \rightarrow a}(x_i) \\
 & = (1-p_0) (1- \prod_{v_i \in V} \mu_{v_i \rightarrow a}(\textit{true}))\\
\end{align*}

After normalization, the sent messages are:
\[
\mu_{a \rightarrow v}^{'}(\textit{true}) = \frac{\mu_{a \rightarrow v}(\textit{true})}{\mu_{a \rightarrow v}(\textit{true}) + \mu_{a \rightarrow v}(\textit{false})}
\]
\[
\mu_{a \rightarrow v}^{'}(\textit{false}) = \frac{\mu_{a \rightarrow v}(\textit{false})}{\mu_{a \rightarrow v}(\textit{true}) + \mu_{a \rightarrow v}(\textit{false})}
\]

For the message from factor $a$ to a random variable $v^* \in V$, we denote $\overline{V} = V \backslash v^*$ and $\overline{X} = X \backslash x^*$, the message is:

\[
    \mu_{a \rightarrow v^*}(x^*) = \sum_{y,\overline{X}} f(y,x^*,\overline{X}) \cdot \mu_{v \rightarrow a}(y) \cdot \prod_{v_i \in \overline{V}} \mu_{v_i \rightarrow a}(x_i)
\]

Let $b = p_0 \cdot \mu_{v\rightarrow a}(\textit{true}) + (1-p_0) \cdot \mu_{v\rightarrow a}(\textit{false})$, we have:
\begin{align*} 
\mu_{a \rightarrow v^*}(\textit{true}) &=\sum_{y,\overline{X}} f(y,\textit{true},\overline{X}) \cdot \mu_{v \rightarrow a}(y) \cdot \prod_{v_i \in \overline{V}} \mu_{v_i \rightarrow a}(x_i)\\
 & = (\mu_{v\rightarrow a}(\textit{true}) - b) \prod_{v_i \in \overline{V}} \mu_{v_i \rightarrow a}(\textit{true}) + b\\
\end{align*}
\begin{align*} 
\mu_{a \rightarrow v^*}(\textit{false}) &=\sum_{y,\overline{X}} f(y,\textit{false},\overline{X}) \cdot \mu_{v \rightarrow a}(y) \cdot \prod_{v_i \in \overline{V}} \mu_{v_i \rightarrow a}(x_i)\\
 & = p_0 \cdot \mu_{v\rightarrow a}(\textit{true}) + (1-p_0) \cdot \mu_{v\rightarrow a}(\textit{false})\\
\end{align*}

After normalization, the sent messages are:
\[
\mu_{a \rightarrow v^*}^{'}(\textit{true}) = \frac{\mu_{a \rightarrow v^*}(\textit{true})}{\mu_{a \rightarrow v^*}(\textit{true}) + \mu_{a \rightarrow v^*}(\textit{false})}
\]
\[
\mu_{a \rightarrow v^*}^{'}(\textit{false}) = \frac{\mu_{a \rightarrow v}(\textit{false})}{\mu_{a \rightarrow v^*}(\textit{true}) + \mu_{a \rightarrow v^*}(\textit{false})}
\]

The details of the deduction of $\mu_{a \rightarrow v}(\textit{true})$, $\mu_{a \rightarrow v}(\textit{false})$, $\mu_{a \rightarrow v^*}(\textit{true})$, and $\mu_{a \rightarrow v^*}(\textit{false})$ can be found in the Appendix.

In summary, by exploiting the structure of our conditional probability distributions, we have simplified the exponential calculation process in the original loopy belief propagation into a linear one.

\section{Implementation for Java}
\label{sec:implementation}
In this section, we describe how to instantiate our approach for Java programs, with a focus on the implementation details about how to collect and analyze the dynamic information of test runs to build the dynamic dependency graph.
In the context of Java programs, a statement is a bytecode instruction which we refer to as an ``instruction'' for short, and a statement instance is referred to as an ``instruction execution''.

The first step is to collect dynamic information from tests. Our approach instruments test runs at the bytecode level and collect traces that are later used to build the dynamic dependency graph. Concretely, such a trace includes a sequence of instruction executions, where each instruction execution includes the ID of the instruction, the type of the instruction, the constants in the instruction (e.g. the branch offset in ``goto''), the values read/written by this instruction, and the source code location corresponding to the instruction. 
In addition, if an exception catch occurs, we will record the current stack trace information in the trace. 
After tracing, we can capture the data and control dependencies on the collected traces by simulating their executions to build the dynamic dependency graph described in ~\defref{def:ddg}. For control dependencies, we also need to apply a static analysis on the whole program.
%


To capture data dependencies, we perform a dynamic analysis on the collected traces, simulating the execution process of tests.
Since we only need to know how the values in each variable and memory address propagate to each other, the data structures we maintain include a call stack and a heap. 
The call stack includes various frames, each of which represents a method invocation. 
A new frame is created and pushed into the stack each time a method is invoked. A frame is destroyed and popped from the stack when its method invocation completes. 
Each frame has its own operand stack and local variable array. 
The heap is a map that contains class instances and arrays. We use the object address and the field name as the key to access the heap. For arrays, we do not distinguish elements in the same array, and abstract an array into a single object to keep our method efficient.
The elements held in the stack and the heap are random variables for each run-time value. In addition, if a run-time value is an object reference or array reference, we will store its actual value (i.e., an address) as the key to access the heap.

When parsing the bytecode sequence, We simulate the operation of instructions on the operand stack and local variable array in the current stack frame, as well as on the heap.
If we hit an exception stack trace, we will adjust the call stack to its consistent structure by excluding stack frames that complete abruptly due to exceptions.
For each instruction, we find the elements it reads from the data structure and store the elements it writes into the data structure. In this process, the statement is evaluated based on the read values to change the written value.
Corresponding data dependencies are transformed into items in the dynamic dependency graph.

\begin{figure}[t]
\footnotesize
\centering
\lstset{language=Java,basicstyle=\tt\footnotesize}
\begin{lstlisting}[frame=single]
public int foo(...) {
    if (condition)
        return a;
    ...
    return b;
}
\end{lstlisting}
\caption{Example Program Demonstrating Control Dependencies.}
\label{fig:dependency}
\end{figure}

Besides data dependencies, we also need to consider control dependencies to precisely model how errors are introduced and propagated.
Unlike modeling data dependencies, we need to consider information from the whole program rather than only from traces.
To decide whether a branching statement controls another statement in execution, one needs to investigate whether the statement would still be executed if the branching statement turns to a different branch other than the one taken in the execution.
Therefore, we need to perform a static analysis of the control flow graph.
Consider the program in \figref{fig:dependency}.
Suppose in a concrete run, the program takes the false branch, and thus the trace is ``if(false){} ... return b''.
The trace does not include the information of the true branch and it remains unclear whether line 5 would be executed if the true branch was taken.
However, by investigating the program, we know that the dependency holds as the method will return at line 3 if the branching statement takes the true branch.
As a result, the branching statement controls all other statements in the method, as its branching status affects all other statements being executed or not.
To precisely model control dependencies, we apply a static analysis~\cite{cooper2001simple} that calculates dominance relations~\cite{dragonbook} between statements on control flow graphs.
Intuitively, a branching statement controls all its subsequent statements in its control flow graph until it reaches statements that post-dominate it.

In order to combine the calculated control dependencies with data dependencies, the call stack frame described above also needs to contain a predicate stack to store the elements corresponding to the calculated values of branching instructions. 
Each time a branching statement is executed, the corresponding element is pushed into the current predicate stack. When the post-dominate statement of the branching statement is reached, the predicate stack will be popped. 
When constructing the dynamic dependency graph, only the predicate value at the top of the predicate stack (if the current position is not controlled by any statement, the predicate stack should be empty) will be treated as the program state value which controls whether this statement is executed. Corresponding control dependencies are transformed into items in the dynamic dependency graph.

With regard to control dependencies, one difficult issue is handling exception control flows. 
To balance precision and soundness, we only consider exceptions that are thrown in the trace.
A fully sound approach would require considering every statement that can throw an exception (e.g., a field access can throw a NullPointerException), which would introduce a lot of spurious dependencies.
At present, our handling of control dependencies involving exceptions is as follows: after hitting an exception stack trace, the captured exception instance is pushed into the current predicate stack. This node will never be popped until the method is finished.
The explanation of this processing is that the exception causes the program to jump to the location where the exception is captured, so the exception instance has a direct control effect on subsequent statements.

\section{Empirical Evaluation \label{sec:evaluation}}
\subsection{Research Questions}
\begin{itemize}
    \item{\bf RQ1: Effectiveness of SmartFL.} How effective is SmartFL compared to other standalone techniques?    
    \item{\bf RQ2: Efficiency of SmartFL.} What is the time cost of SmartFL compared to other techniques?
    \item{\bf RQ3: Effectiveness of Different Components.} What is the contribution of each component to the overall effectiveness of SmartFL?
    \item {\bf RQ4: Influence of Different Factor Values.} 
    To what extent do moderate, very low, and very high factor values in both sensitive and insensitive statements affect the results?
    \item {\bf RQ5: Combining with other Techniques.} Can SmartFL improve the effectiveness of combination methods?
    
\end{itemize}
\subsection{Benchmark and Measurements}
\begin{table}[t]
  \caption{Projects from Defects4J dataset, version 2.0.0.}
  \label{tab:dataset}
  \begin{center}
    \begin{tabular}{l|llll}
    \toprule
    Project & Faults & LoC & ATests & CTests\\
    \midrule
    Chart & 26 & 203.0k & 1818 & 38\\
    Cli & 39 & 5.7k & 262 & 42\\
    Closure & 174 & 138.8 k & 7027 & 18\\
    Codec & 18 & 10.9k & 440 & 32\\
    Collections & 4 & 67.0k & 15582 & 34\\
    Compress & 47 & 31.0k & 432 & 40\\
    Csv & 16 & 3.1k & 180 & 39\\
    Gson & 18 & 14.0k & 988 & 33\\
    JacksonCore & 26 & 34.4k & 356 & 41\\
    JacksonDatabind & 112 & 95.8k & 1610 & 13\\
    JacksonXml & 6 & 7.6k & 152 & 40\\
    Jsoup & 93 & 15.0k & 454 & 18\\
    JxPath & 22 & 29.2k & 305 & 12\\
    Lang & 64 & 52.3k & 1815 & 30\\
    Math & 106 & 116.2k & 3343 & 29\\
    Mockito & 38 & 18.8k & 1156 & 5\\
    Time & 26 & 67.7k & 3802 & 21\\
    Total & 835 & 76.3k & 2604 & 24\\
  \bottomrule
  \end{tabular}
  \end{center}
  {\small `Faults' denotes the number of defective versions of the project, `LoC' denotes the average lines of code of each project 
  , `ATests' denotes the average test numbers of each project, and `CTests' denotes the average number of chosen tests after reducing redundant tests
  .}
\end{table}

We take the projects from Defects4J~\cite{just2014defects4j} version 2.0 as our benchmark suite. Defects4J 2.0 includes 835 faults from 17 projects, as shown in \tabref{tab:dataset}, where ``ATests'' denotes the average test numbers of each project, and ``CTests'' denotes the average number of chosen tests after reducing redundant tests. Notice that since our approach does not rely on machine learning, we do not need to use this dataset as a training set.

Our evaluating metric is top-k where $k$ is 1, 3, 5, or 10. Top-k counts the number of faults that are successfully located within the top k entries of the ranked suspicious candidate list. An existing study ~\cite{topk_parnin2011automated} suggested that developers would only check a few entries in the ranked list, which is consistent with the top-k metric. 
We manually mark all fault locations and follow the tie-break rules provided in a previous study~\cite{CombineFL}.

Regarding the granularity of fault localization, we choose both statement-level and method-level granularity, the two most frequently used levels. 
As for method-level evaluation, we calculate the maximum suspicious score of the statements in a method as the suspicious score of the method.

\subsection{Experiment Setup\label{setup}}
\subsubsection{Setting}
We have implemented our approach for Java using the instrumentation framework Javassist\footnote{\url{https://github.com/jboss-javassist/javassist}}. 
As described in \secref{sec:selecting}, we select up to 50 test methods for tracing and limit the maximum number of statements to less than 1.2 million for each trace. Upon building the graph, we limit the maximum number of statements to less than 1 million for all compressed traces. 
Please note that this selection only applies to our approach but not any other baseline approach. 


\subsubsection{RQ1: Effectiveness of SmartFL} 
SmartFL is a standalone fault localization technique that does not rely on other techniques.
To test the effectiveness of SmartFL, we compare the result of SmartFL with other standalone techniques. We choose SBFL to represent coverage-based approaches and MBFL to represent the approaches modeling semantics. We do not compare it with angelic debugging because there is no implementation scalable to large programs as far as we know. According to existing research~\cite{CombineFL}, we select Ochiai~\cite{sbfl_abreu2007accuracy} and DStar~\cite{sbfl_wong2013dstar} from SBFL, Metallaxis~\cite{mbfl_papadakis2015metallaxis} and MUSE~\cite{mbfl_moon2014ask} from MBFL for comparison. Pearson et al. \cite{pearson2016evaluating} studied the performance of SBFL and MBFL on Defects4J, and our experiments reuse their implementation. 
In addition, we also choose a set of state-of-the-art fault localization approaches as baselines, and the results of these baselines are obtained from their papers directly. 
CAN~\cite{can} leverages graph neural networks to analyze and combine the failure context for fault localization. UNITE~\cite{unite} optimizes coverage information based on the frequency of each statement appearing in each test.
GRACE~\cite{grace} leverages graph-based representation learning to embed both the syntax of the program and the coverage information. LEAM~\cite{leam} is a state-of-the-art MBFL approach, that leverages DL-based mutation techniques.
For CAN and UNITE, we cannot directly compare the results with them because there was no artifact that could be used to reproduce the experimental results in these papers.
Therefore, we can only take an indirect comparison based on the data presented in their papers. Concretely, their papers report the ratios of top-k, and their benchmarks include four Defects4J projects: Lang, Math, Chart, and Time. Although their benchmarks also include other projects outside of Defects4J, we can directly determine the upper limits of the numbers of top-k in these four projects by multiplying the numbers of cases in each benchmark by the ratios of top-k, and compare them with SmartFL.
For example, CAN shows that its benchmark includes 318 cases, of which 224 are from Defects4J and the top-1 ratio is 4.62\%, so we can assume that the top-1 result must be less than $318 * 4.62\% = 15$ for cases from Defects4J.
For GRACE and LEAM, they only report method-level results, thus we can only compare method-level results with them. GRACE conducts experiments on all projects in Defects4J 2.0, while LEAM reports method-level results on 4 projects in Defects4J 1.0, including Lang, Math, Chart, and Time.
As a result, we separately compare SmartFL with them.
We will discuss combination techniques in \textit{RQ5}.

\subsubsection{ RQ2: Efficiency of SmartFL \label{sec:efficiency}}
The time cost in fault localization is also important, as an approach that takes only a few minutes is much more convenient than an approach that takes several hours. In this RQ, 
we compare the run-time cost of SmartFL with the four baseline approaches described in RQ1. 
As \figref{fig:framework} shows, SmartFL consists of three steps: (a) profiling (coarse-grained instrumentation to get method-level coverage, described in \secref{sec:selecting}), (b) tracing (getting fine-grained traces of selected tests), and (c) modeling (building the probabilistic graph and probabilistic inference). In the profiling step, we use the default test instruction of Defects4J to execute the entire test suite in a single-threaded.
In the tracing step, each test can run in parallel except for project Time and Closure because the tests of Time and Closure do not support parallel execution. 
As a result, we run tests in parallel with 16 threads for the other three projects and run tests with a single thread for Time and Closure. In the modeling step, we run modeling and probabilistic inference in a single thread. For SBFL and MBFL methods, we use the implementation from \cite{pearson2016evaluating} and collect the time-consumption data.

\subsubsection{RQ3: Effectiveness of Different Components \label{component}} We design six ablation studies to evaluate the effectiveness of each component. 
\begin{enumerate}
    \item In the first study, we use the original version of SmartFL from \cite{DBLP:conf/icse/ZengWY0Z022} to compare the improvements of the new version compared to the old version. 
    \item In the second study, we discard the adaptive folding technology described in \secref{sec:partial} to see the impact of the adaptive folding technology.
    \item In the third study, we discard the handling of exception control flow described in \secref{sec:implementation} to see the impact of modeling exception control flow.
    \item In the fourth study, we discard the modeling of ``virtual call edge'' described in \secref{sec:partial} to see the impact of this modeling.
    \item In the fifth study, we discard loop compression described in \secref{sec:compress} to show the impact of loop compression.
    \item In the sixth study, we discard the inference algorithm optimization described in \secref{sec:optimization} to show the impact of this optimization.
    \item In the seventh study, we discard the test reduction described in \secref{sec:selecting} to show the impact of this optimization.
\end{enumerate}
In the second and sixth studies, due to the lack of inference algorithm optimization, the modeling part cannot be completed within the valid time in some cases. Therefore, we introduced a 20-minute time limit for the modeling part of each case. In the seventh study, we only conducted the experiment on project Lang for scalability issues. Specifically, we only exclude the passing tests that have no coverage intersection with the failing tests and consider all other tests in modeling.
When calculating the top-k results and average time consumption, only cases that have not timed out will be considered. At the same time, we will give the number of timed-out cases in each study. 

For other techniques like partial tracing, they are a fundamental part of our framework for efficiency. Therefore, we do not design corresponding ablation studies.

%
%
%


\subsubsection{RQ4: Influence of Different Factor Values} 
Our approach contains a parameter about the probability $P(Result = 1 | Parents = 0)$. We assign 0.5 to the factor values of insensitive operations and 0.01 to those of sensitive operations in our default approach.
In this RQ we evaluate the performance of other possible values. We evaluate 5 $p_0$ values for insensitive operations: $0.3, 0.4, 0.5, 0.6, 0.7$, and 5 $p_0$ values for sensitive operations: $0.001, 0.005, 0.01, 0.05, 0.1$. 


\subsubsection{RQ5: Combining with other Techniques} 
CombineFL~\cite{CombineFL} is one of the state-of-the-art combination-based fault localization techniques on the statement level. CombineFL combines different fault localization methods, including history-based, stack trace-based, IR-based, slicing, SBFL, predicates switching, and MBFL. CombineFL uses rankSVM~\cite{DBLP:conf/sdm/KuoLL14} to combine the results of multiple methods so that our method can be easily combined with other methods.
TRANSFER-FL~\cite{DBLP:conf/icse/MengW00022} is another fault localization method that combines multiple types of information. TRANSFER-FL designs BiLSTM-based classifiers to learn deep semantic features of statements from open-source bug datasets and leverages the semantic-based, spectrum-based, and mutation-based features for effective fault localization by a multi-layer perceptron. 
TRANSFER-FL is also one of the state-of-the-art fault localization techniques on the statement level, as CombineFL and TRANSFER-FL are relatively close in performance on Defects4J 1.0. 

We integrated the results of SmartFL into CombineFL and compared them with the results of CombineFL and TRANSFER-FL to verify the ability of SmartFL to combine with other methods. We combine SmartFL with other methods under the framework of CombineFL on 357 cases from Defects4J 1.0 because other methods used by CombineFL only include the results of these cases. CombineFL requires a suspiciousness score for each standalone technique to perform learning.
In SmartFL, the suspiciousness score of an $i_{th}-ranked$ statement is defined as follow:
$Suspiciousness(i) = \frac{n-i+1}{n}$, where $n$ is the number of all suspicious candidates. We set up a simple heuristic screening strategy to select high-quality results, that is, only use cases with a size of less than 300 statements in the result file to join the training, while in other cases we use the original data.

 
\subsection{Experiment Results \label{sec:results}}
\subsubsection{RQ1: Effectiveness of SmartFL}
\begin{table}[t]
  \caption{Statement-level Performance. The result continues in \tabref{tab:stmt2} }
  \label{tab:stmt}
  \begin{center}
    \begin{tabular}{p{1.2cm}|p{1.2cm}|p{1cm}p{1cm}p{1cm}p{1cm}}
    \toprule
    Project & Technique & Top-1 & Top-3 & Top-5 & Top-10\\
    \midrule
    \multirow{5}{*}{Lang} & Ochiai & 2(3\%) & 21(33\%) & 28(44\%) & 37(58\%) \\
    & DStar & 1(2\%) & 14(22\%) & 19(30\%) & 29(61\%)\\
    & Metallaxis & 8(13\%) & 24(38\%) & 33(52\%) & 42(66\%)\\
    & MUSE & 8(13\%) & 13(20\%) & 15(23\%) & 15(23\%)\\
    \rowcolor{tablegray}
    \cellcolor{white}& SmartFL & \textbf{20(31\%)} & \textbf{35(55\%)} & \textbf{39(61\%)} & \textbf{45(71\%)}\\
    \midrule
    \multirow{5}{*}{Math} & Ochiai & 11(10\%) & \textbf{30(28\%)} & \textbf{41(39\%)} & \textbf{61(57\%)} \\
    & DStar & 10(9\%) & 21(20\%) & 28(26\%) & 46(43\%)\\
    & Metallaxis & 10(9\%) & 28(26\%) & 39(37\%) & 43(41\%)\\
    & MUSE & 7(7\%) & 15(14\%) & 22(20\%) & 25(24\%)\\
    \rowcolor{tablegray}
    \cellcolor{white}& SmartFL & \textbf{16(15\%)} & \textbf{30(28\%)} & \textbf{41(39\%)} & 43(41\%)\\
    \midrule
    \multirow{5}{*}{Chart} & Ochiai & 1(4\%) & 6(23\%) & 8(31\%) & 15(58\%) \\
    & DStar & 1(4\%) & 5(19\%) & 7(27\%) & 13(50\%)\\
    & Metallaxis & 0(0\%) & 6(23\%) & 6(23\%) & 11(42\%)\\
    & MUSE & 2(8\%) & 4(15\%) & 5(19\%) & 5(19\%)\\
    \rowcolor{tablegray}
    \cellcolor{white}& SmartFL & \textbf{5(19\%)} & \textbf{14(54\%)} & \textbf{16(62\%)} & \textbf{19(73\%)}\\
    \midrule
    \multirow{5}{*}{Time} & Ochiai & 1(4\%) & 7(27\%) & \textbf{9(35\%)} & 11(42\%) \\
    & DStar & 3(12\%) & 7(27\%) & \textbf{9(35\%)} & \textbf{12(46\%)}\\
    & Metallaxis & \textbf{4(15\%)} & \textbf{8(31\%)} & \textbf{9(35\%)} & 11(42\%)\\
    & MUSE & 0(0\%) & 1(4\%) & 2(8\%) & 4(15\%)\\
    \rowcolor{tablegray}
    \cellcolor{white}& SmartFL & \textbf{4(15\%)} & 7(27\%) & 8(31\%) & 11(42\%)\\
        \midrule
    \multirow{5}{*}{Closure} & 
    Ochiai & \textbf{7(4\%)} & \textbf{16(9\%)} & \textbf{24(14\%)} & \textbf{40(23\%)} \\
    & DStar & 3(2\%) & 14(8\%) & 20(11\%) & 34(20\%)\\
    & Metallaxis & 1(1\%) & 7(4\%) & 10(6\%) & 13(7\%)\\
    & MUSE & 6(3\%) & 8(5\%) & 10(6\%) & 14(8\%)\\
    \rowcolor{tablegray}
    \cellcolor{white}& SmartFL & 6(3\%) & 8(5\%) & 10(6\%) & 14(8\%)\\
        \midrule
    \multirow{5}{*}{Mockito} & 
    Ochiai & 3(8\%) & 8(21\%) & 8(21\%) & 8(21\%) \\
    & DStar & 2(5\%) & 7(18\%) & 7(18\%) & 7(18\%)\\
    & Metallaxis & 3(8\%) & 6(16\%) & 7(18\%) & 11(29\%)\\
    & MUSE & 2(5\%) & 7(18\%) & 9(24\%) & 9(24\%)\\
    \rowcolor{tablegray}
    \cellcolor{white}& SmartFL & \textbf{10(26\%)} & \textbf{13(34\%)} & \textbf{14(37\%)} & \textbf{16(42\%)}\\
        \midrule
    \multirow{5}{*}{Cli} & 
    Ochiai & 6(15\%) & \textbf{12(31\%)} & \textbf{16(41\%)} & \textbf{20(51\%)} \\
    & DStar & 5(13\%) & 7(18\%) & 12(31\%) & 15(38\%)\\
    & Metallaxis & 3(8\%) & 7(18\%) & 9(23\%) & 10(26\%)\\
    & MUSE & 2(5\%) & 4(10\%) & 4(10\%) & 6(15\%)\\
    \rowcolor{tablegray}
    \cellcolor{white}& SmartFL & \textbf{7(18\%)} & 9(23\%) & 11(28\%) & 14(36\%)\\
        \midrule
    \multirow{5}{*}{Codec} & 
    Ochiai & 2(11\%) & 6(33\%) & 6(33\%) & 9(50\%) \\
    & DStar & 3(17\%) & 6(33\%) & 6(33\%) & 8(44\%)\\
    & Metallaxis & 1(6\%) & 2(11\%) & 211\%) & 5(28\%)\\
    & MUSE & 1(6\%) & 3(17\%) & 3(17\%) & 5(28\%)\\
    \rowcolor{tablegray}
    \cellcolor{white}& SmartFL & \textbf{5(28\%)} & \textbf{7(39\%)} & \textbf{8(44\%)} & \textbf{10(56\%)}\\
        \midrule
    \multirow{5}{*}{Collections} & 
    Ochiai & 0(0\%) & \textbf{1(25\%)} & \textbf{1(25\%)} & \textbf{1(25\%)} \\
    & DStar & 0(0\%) & 0(0\%) & 0(0\%) & 0(0\%)\\
    & Metallaxis & \textbf{1(25\%)} & \textbf{1(25\%)} & \textbf{1(25\%)} & \textbf{1(25\%)} \\
    & MUSE & \textbf{1(25\%)} & \textbf{1(25\%)} & \textbf{1(25\%)} & \textbf{1(25\%)}\\
    \rowcolor{tablegray}
    \cellcolor{white}& SmartFL & \textbf{1(25\%)} & \textbf{1(25\%)} & \textbf{1(25\%)} & \textbf{1(25\%)}\\
  \bottomrule
  \end{tabular}
  \end{center}
\end{table}
\begin{table}[t]
  \caption{Statement-level Performance. Continued from  \tabref{tab:stmt} }
  \label{tab:stmt2}
  \begin{center}
    \begin{tabular}{p{1cm}|p{1.2cm}|p{1cm}p{1cm}p{1cm}p{1cm}}
    \toprule
    Project & Technique & Top-1 & Top-3 & Top-5 & Top-10\\
        \midrule
    \multirow{5}{*}{Compress} & 
    Ochiai & 5(11\%) & 14(30\%) & 16(34\%) & 18(38\%) \\
    & DStar & 5(11\%) & 14(30\%) & 15(32\%) & 18(38\%) \\
    & Metallaxis & 2(4\%) & 8(17\%) & 11(23\%) & 17(36\%)\\
    & MUSE & 0(0\%) & 7(15\%) & 9(19\%) & 13(28\%)\\
    \rowcolor{tablegray}
    \cellcolor{white}& SmartFL & \textbf{7(15\%)} & \textbf{15(32\%)} & \textbf{19(40\%)} & \textbf{24(51\%)}\\
        \midrule
    \multirow{5}{*}{Csv} & 
    Ochiai & \textbf{4(25\%)} & \textbf{6(38\%)} & \textbf{7(22\%)} & \textbf{8(50\%)} \\
    & DStar & \textbf{4(25\%)} & \textbf{6(38\%)} & \textbf{7(22\%)} & \textbf{8(50\%)} \\
    & Metallaxis & 2(12\%) & 4(25\%) & 4(25\%) & 6(38\%)\\
    & MUSE & 1(6\%) & 2(12\%) & 2(12\%) & 3(19\%)\\
    \rowcolor{tablegray}
    \cellcolor{white}& SmartFL & \textbf{4(25\%)} & 5(31\%) & 6(38\%) & 6(38\%)\\
        \midrule
    \multirow{5}{*}{Gson} & 
    Ochiai & 2(11\%) & 4(22\%) & 6(33\%) & 8(44\%) \\
    & DStar & 2(11\%) & 4(22\%) & 6(33\%) & 8(44\%) \\
    & Metallaxis & 5(28\%) & 5(28\%) & 5(28\%) & 5(28\%)\\
    & MUSE & 1(6\%) & 2(11\%) & 2(22\%) & 2(22\%)\\
    \rowcolor{tablegray}
    \cellcolor{white}& SmartFL & \textbf{6(33\%)} & \textbf{9(50\%)} & \textbf{9(50\%)} & \textbf{9(50\%)}\\
        \midrule
    \multirow{5}{*}{JCore} & 
    Ochiai & 3(12\%) & 7(27\%) &7(27\%) &7(27\%) \\
    & DStar &1(4\%) & 7(27\%) &7(27\%) &8(31\%) \\
    & Metallaxis & 4(15\%) & 6(23\%) & 8(31\%) & 8(31\%)\\
    & MUSE & 2(8\%) & 2(8\%) & 2(8\%) & 3(12\%)\\
    \rowcolor{tablegray}
    \cellcolor{white}& SmartFL & \textbf{6(23\%)} & \textbf{8(31\%)} & \textbf{9(35\%)} & \textbf{11(42\%)}\\
        \midrule
    \multirow{5}{*}{JDatabind} & 
    Ochiai & 0(0\%) & 0(0\%) & 0(0\%) & 0(0\%) \\
    & DStar & 0(0\%) & 0(0\%) & 0(0\%) & 0(0\%) \\
    & Metallaxis & 1(1\%) & 5(4\%) & 6(5\%) & 10(9\%)\\
    & MUSE & 1(1\%) & 2(2\%) & 3(3\%) & 5(4\%)\\
    \rowcolor{tablegray}
    \cellcolor{white}& SmartFL & \textbf{6(5\%)} & \textbf{12(11\%)} & \textbf{13(12\%)} & \textbf{18(16\%)}\\
        \midrule
    \multirow{5}{*}{JXml} & 
    Ochiai & 0(0\%) & 0(0\%) & 0(0\%) & 0(0\%) \\
    & DStar & 0(0\%) & 0(0\%) & 0(0\%) & 0(0\%) \\
    & Metallaxis & 0(0\%) & \textbf{2(33\%)} & \textbf{2(33\%)} & \textbf{2(33\%)}\\
    & MUSE & 0(0\%) & 0(0\%) & 0(0\%) & 0(0\%)\\
    \rowcolor{tablegray}
    \cellcolor{white}& SmartFL & 0(0\%) & 0(0\%) & 0(0\%) & 0(0\%)\\
        \midrule
    \multirow{5}{*}{Jsoup} & 
    Ochiai & 3(3\%) & 4(4\%) & 5(5\%) & 11(12\%) \\
    & DStar & 0(0\%) & 2(2\%) & 3(3\%) & 8(9\%)\\
    & Metallaxis & 6(6\%) & 13(14\%) & 14(15\%) & 20(22\%)\\
    & MUSE & 0(0\%) & 1(1\%) & 3(3\%) & 8(9\%)\\
    \rowcolor{tablegray}
    \cellcolor{white}& SmartFL & \textbf{11(12\%)} & \textbf{23(25\%)} & \textbf{28(30\%)} & \textbf{32(34\%)}\\
        \midrule
    \multirow{5}{*}{JxPath} & 
    Ochiai & 0(0\%) & 0(0\%) & 0(0\%) & 0(0\%) \\
    & DStar & 0(0\%) & 0(0\%) & 0(0\%) & 0(0\%) \\
    & Metallaxis & 0(0\%) &0(0\%) & 0(0\%) & 4(18\%)\\
    & MUSE & 2(9\%) & 3(14\%) & 3(14\%) & 3(14\%)\\
    \rowcolor{tablegray}
    \cellcolor{white}& SmartFL & \textbf{1(5\%)} & \textbf{4(18\%)} & \textbf{6(27\%)} & \textbf{6(27\%)}\\
    \midrule
    \multirow{5}{*}{Total} & 
    Ochiai & 50(6\%) & 142(17\%) & 182(22\%) & 254(30\%) \\
    & DStar & 42(5\%) & 114(14\%) & 146(17\%) & 214(26\%)\\
    & Metallaxis & 51(6\%) & 132(16\%) & 166(20\%) & 219(26\%)\\
    & MUSE & 36(4\%) & 75(9\%) & 95(11\%) & 121(14\%)\\
    \rowcolor{tablegray}
    \cellcolor{white}& SmartFL & \textbf{115(14\%)} & \textbf{200(24\%)} & \textbf{238(29\%)} & \textbf{279(33\%)}\\
  \bottomrule
  \end{tabular}
  \end{center}
{\small `JCore' denotes `JacksonCore', and the same for `JDatabind' and `JXml'. }
\end{table}

\tabref{tab:stmt} and \tabref{tab:stmt2} show the numbers and percentages of faults localized by our approach, Ochiai, DStar, Metallaxis, and MUSE at the statement level.
Here, we describe that a fault is successfully localized by an approach if the actual fault position can be found in the top k program elements returned by the approach.
We display the results with different values of k in the top-k metric.
The best results under each category are in bold fonts. 
On all the 835 faults, SmartFL performs best among all values of k. 
At top-1, 3, 5, and 10, SmartFL improves 115\%, 41\%, 31\%, and 10\% over the second-best approach, respectively. 
Regarding the details of each project, our approach outperforms the baselines on most of the benchmarks.
For the few exceptions where Ochiai performs better, our approach fails to capture part of the dynamic dependencies. 
If the dependency is missed, our approach will not be able to capture the relationship between the faulty statement and the failure of the test, thus incorrectly ignoring the faulty statement.
For instance, our approach does not model the data dependency introduced through IO (i.e. writing to a file and then reading from it), which we plan to address in the future.

\begin{table}[t]
  \caption{Method-level Performance}
  \label{tab:method}
  \begin{center}
    \begin{tabular}{c|c|cccc}
    \toprule
    Technique & Top-1 & Top-3 & Top-5 & Top-10\\
    \midrule
    Ochiai & 167(20\%) & 305(37\%) & 351(42\%) & 398(48\%) \\
    DStar & 157(19\%) & 274(33\%) & 316(38\%) & 371(44\%)\\
    Metallaxis & 143(17\%) & 261(31\%) & 301(36\%) & 351(42\%)\\
    MUSE & 90(11\%) & 158(19\%) & 188(23\%) & 220(26\%)\\
    GRACE & \textbf{280(34\%)} & \textbf{382(46\%)} & \textbf{438(52\%)} & $\backslash$ \\
    \rowcolor{tablegray}
    SmartFL & 213(26\%) & 326(39\%) & 372(45\%) & 424(51\%)\\
  \bottomrule
  \end{tabular}
  \end{center}
\end{table}

\tabref{tab:method} shows the performance of each approach at the method level. 
It also includes the method-level results produced by GRACE which is designed specifically for method-level fault localization.
The GRACE paper conducts experiments on all projects in Defects4J 2.0 and we directly use the results from it.
Similar to the statement-level results, SmartFL outperforms SBFL methods and MBFL methods among all values of k. However, SmartFL is not as good as GRACE at the method level. We think this result is reasonable because the design of SmartFL is mainly aimed at statement-level fault localization rather than method-level as our modeling is based on the execution process of each statement.
In contrast, the method-level results are not our focus, because the direct output of SmartFL is the statements sorted by faulty probability, and the method-level results are directly obtained by statement-level results.

\begin{table}[t]
  \caption{Comparing SmartFL with LEAM}
  \label{tab:leam}
  \begin{center}
    \begin{tabular}{c|c|cccc}
    \toprule
    Technique & Top-1 & Top-3 & Top-5 \\
    \midrule
    LEAM-Metallaxis & 118(53\%) & 182(81\%) & 188(84\%) \\
    LEAM-MUSE & \textbf{126(56\%)} & \textbf{181(81\%)} & \textbf{189(84\%)}\\
    \rowcolor{tablegray}
    SmartFL & 91(41\%) & 131(58\%) & 149(67\%) \\
  \bottomrule
  \end{tabular}
  \end{center}
\end{table}

We also compare our approach with LEAM at the method level, which is another approach designed specifically for method-level fault localization.
As its paper reports method-level results on only 4 projects in Defects4J 1.0, including Lang, Math, Chart, and Time, we put the comparison results separately in \tabref{tab:leam}.
As \tabref{tab:leam} shows, similar to the comparison with GRACE, LEAM is also more effective than SmartFL at the method level.

\begin{table}[t]
  \caption{Sign Test Result}
  \label{tab:sign}
  \begin{center}
    \begin{tabular}{c|c|c|c|c|c}
    \toprule
    Level & Method &	POS &	NEG &	P-value & Effect Size\\
    \midrule
\multirow{4}{*}{Statement} & vs. Ochiai & 209 & 134 & 3.03e-5 & 0.27\\
& vs. DStar &	222 &	122 &	3.84e-8 & 0.51\\
& vs. Metallaxis &	218 &	136 &	7.69e-6 & 0.42\\
& vs. Muse &	248 &	90 &	1.59e-18 & 1.09\\
\midrule
\multirow{4}{*}{Method} & vs. Ochiai &	227 &	189 &	0.0348 & 0.10\\
& vs. DStar &	243 &	180 &	0.0012 & 0.21 \\
& vs. Metallaxis &	235 &	171 &	8.67e-4 & 0.27\\
& vs. Muse &	312 &	124 &	4.47e-20 &0.99\\
  \bottomrule
  \end{tabular}
  \end{center}
\end{table}

We perform a sign test on each pair of techniques considering faults where at least one technique has a top-10 result on it, and the result is shown in \tabref{tab:sign}. 
GRACE and LEAM are excluded from the comparison as we only take their results directly and do not manually reproduce them.
We confine the test to these faults because a difference between rank 100 and 1000 would not make a great difference in actual use cases. In addition, we use the standardized difference between two means as the effect size~\cite{sullivan2012using}. In our setting, if a method's result on a particular example is not in top-10, it will be set to 15 for calculating the effect size because we care most about  top-10 cases as in the sign test, and should give a reasonable effect size to the cases that are not in top-10.
The result implies that our approach significantly outperforms other methods, as all p-values are less than the significance level of 0.05.
Still, the negative cases in the sign test show that our approach could be complementary to others, which is further discussed in RQ5. The effect sizes show that our approach is more suitable for statement-level fault localization as the effect size is small against Ochiai, medium against Dstar and Metallaxis, and large against Muse. On the other hand, for method-level fault localization, the effect size is small against Ochiai, Dstar, Metallaxis, and large against Muse.

\begin{table}[t]
  \caption{Comparing SmartFL with CAN and UNITE}
  \label{tab:can}
  \begin{center}
    \begin{tabular}{c|c|cccc}
    \toprule
    Technique & Top-1 & Top-3 & Top-5 \\
    \midrule
    CAN & $\leq 15(7\%)$ & $\leq 64(28\%)$ & $\leq93(41\%)$\\
    UNITE & $\leq 26(12\%)$ & $\leq 75(33\%)$ &  $\leq100(45\%)$ \\
    \rowcolor{tablegray}
    SmartFL & \textbf{47(21\%)} & \textbf{88(39\%)} & \textbf{103(46\%)} \\
  \bottomrule
  \end{tabular}
  \end{center}
\end{table}

Finally, we compare our approach with CAN and UNITE at the statement level. As we discussed in \secref{setup}, we directly compare the upper limit of their results based on their papers with SmartFL.
\tabref{tab:can} shows the results comparing SmartFL with CAN and UNITE. Notice that the results of SmartFL are real for the 224 cases from project Lang, Math, Chart, and Time, and the top-k numbers of CAN and UNITE are strictly smaller than SmartFL's in the table. This result shows that SmartFL performs better than CAN and UNITE.

\begin{Summary}{}{firstsummary}
SmartFL focuses on statement-level fault localization and SmartFL performs better than all other techniques at the statement level. At method level, SmartFL significantly outperforms SBFL methods and MBFL methods, but GRACE and LEAM outperform SmartFL. 
In summary, the above results demonstrate the effectiveness of SmartFL.
\end{Summary}




\subsubsection{RQ2: Efficiency of SmartFL}

\begin{table}[t]
  \caption{Average Time Consumption of each Technique (in seconds)}
  \label{tab:time}
  \begin{center}
    \begin{tabular}{l|l|l|l|l|a}
    \toprule
    \multirow{2}{*}{SBFL} & \multirow{2}{*}{MBFL} &
    \multicolumn{4}{c}{SmartFL}\\ \cmidrule{3-6} 
    & &(a) & (b) & (c) & total\\
    \midrule
    413 & 46749 & 41  & 126 & 37 & 205\\

  \bottomrule
  \end{tabular}
  \end{center}
  {\small `SmartFL-(a), (b), and (c)' respectively denote the three steps introduced in \secref{setup}. SmartFL-total denotes the sum of them.}
\end{table}

\tabref{tab:time} shows the time costs of all techniques.
The average time of SmartFL is 205 seconds, which is 50\% of the average time of SBFL methods and an order of magnitude smaller than MBFL methods, while SmartFL is also more accurate.
Although SmartFL does more detailed tracing and performs probabilistic modeling and inference, the results show that it is faster than SBFL methods.
This is mainly because SmartFL will not instrument all tests, but will perform method-level profiling and then only select a part of the tests for tracing according to the method described in \secref{sec:selecting}, while SBFL methods will execute all tests under instrumentation. 
The comparison between columns ``SBFL methods'' and ``SmartFL-(a)'' shows that SmartFL’s method-level profiling is much faster than SBFL’s statement-level profiling. 
For MBFL methods, the average time consumption is about 13 hours, which is much higher than SmartFL.
\begin{Summary}{}{}
The time consumption of SmartFL is half of SBFL methods and 0.5\% of  MBFL methods. These results demonstrate the efficiency of SmartFL.
\end{Summary}

\subsubsection{RQ3: Effectiveness of Different Components}
We perform several ablation studies to evaluate the effect of different components of our approach.
\tabref{tab:ablation} shows the results of the first six experiments. First, let us analyze the impact of the ablation setups shown in the first few columns of the table on the effectiveness of SmartFL.
The first experiment directly compares SmartFL with the original version. Since the original SmartFL timed out on several cases, we also discuss the results of SmartFL after excluding those cases, which are 106, 180, 213, and 249 at top-1, 3, 5, and 10. These two comparisons strictly demonstrate SmartFL has been greatly improved compared to the original version because of the newly introduced components and better implementation.
The second experiment shows that adaptive folding has an important influence on the effectiveness of SmartFL, as it ensures the integrity of the failing test traces. The third experiment shows that the modeling of exception control flow also has an important influence on the effectiveness of SmartFL because it correctly maintains the stack frame while parsing the bytecode sequence and captures the necessary control dependencies. The fourth experiment demonstrates the effect of ``virtual call edge'', which shows that in some cases the lack of construction of the calling edge of the callback function will reduce the effectiveness of SmartFL. The fifth experiment shows that discarding loop compression slightly reduces the effectiveness of SmartFL. The sixth experiment shows that discarding inference optimization will also reduce the effectiveness of SmartFL, as there are some cases where the modeling and inference process cannot be completed without this optimization. 

For the seventh experiment, the results of SmartFL on project Lang after turning off test reduction are 20, 35, 38, and 45 at top-1, 3, 5, and 10, while the original results are 20, 35, 39, and 45, respectively. The results show that there are no significant differences in terms of ranking effectiveness with and without test reduction. 
As a result, our approach of test reduction effecitively increases the scalability of SmartFL without significantly hurting its ranking effectiveness.

\begin{table}[t]
  \caption{Effect of Different Components}
  \label{tab:ablation}
  \begin{center}
    \begin{tabular}{l|llll}
    \toprule
    Technique &  Top-1 & Top-3 & Top-5 & Top-10 \\
    \midrule
    SmartFL& 115 & 200 & 238 & 279\\
    \midrule
    original SmartFL \cite{DBLP:conf/icse/ZengWY0Z022} & 97(-18) & 158(-42) & 188(-50) & 221(-58)\\
    w/o adaptive folding & 103(-12) & 182(-18) & 218(-20) & 257(-22)\\
    w/o exception handling & 103(-12) & 179(-21) & 209(-29) & 244(-35)\\
    w/o virtual call edge & 105(-10) & 185(-15) & 224(-14) & 264(-15)\\
    w/o loop compression & 113(-2) & 194(-6) & 228(-10) & 277(-2)\\
    w/o inference optimization & 102(-13) & 188(-12) & 226(-12) & 264(-15)\\
  \bottomrule
  \end{tabular}
  \end{center}
\end{table}

\begin{table}[t]
  \caption{Time Consumption of Different Components}
  \label{tab:ablation_time}
  \begin{center}
    \begin{tabular}{l|l|l}
    \toprule
    Technique & Time Cost & Timed Out\\
    \midrule
    SmartFL & 200s & 1\\
    \midrule
    original SmartFL & 392s & 242\\
    w/o adaptive folding & 200s & 1\\
    w/o exception handling & 199s & 1\\
    w/o virtual call edge & 201s & 1\\
    w/o loop compression & 205s & 1\\
    w/o inference optimization & 365s & 153\\
  \bottomrule
  \end{tabular}
  \end{center}
\end{table}

\tabref{tab:ablation_time} shows the average time cost and timed-out cases in each experiment. 
We choose the same timed-out settings according to the previous conference paper.
The first experiment shows the efficiency of SmartFL has been greatly improved compared to the original version and there are 242 cases where inference cannot be completed due to lack of inference optimization. The fifth experiment shows that loop compression can slightly improve the efficiency of SmartFL.  The last experiment shows that SmartFL will almost double the average time cost without the inference optimization and will be unable to complete inference on 153 cases. Other experiments show the techniques studied do not affect the efficiency.

\begin{Summary}{}{}
The above results show that each ablation study proves the effectiveness of the corresponding technology in SmartFL. In addition, SmartFL has significantly improved compared with the original version.
\end{Summary}




\subsubsection{RQ4: Influence of Different Factor Values.}
\begin{table}[t]
  \caption{Effect of Different Factor Values}
  \label{tab:parameter}
  \begin{center}
    \begin{tabular}{c|c|cccc}
    \toprule
Parameter Type & Value & Top-1 & Top-3 & Top-5 & Top-10 \\
 \midrule
 \multirow{9}{2.8cm} {Moderate probabilities} 
     & 0.1 & 122 & 203 & 249 & 297 \\
     & 0.2 & 111 & 193 & 237 & 273 \\
    & 0.3 & 117 & 203 & 242 & 286 \\
    & 0.4 & 115 & 203 & 239 & 280 \\
    & 0.5 & 115 & 200 & 238 & 279 \\
    & 0.6 & 114 & 205 & 244 & 284 \\
    & 0.7 & 117 & 208 & 250 & 290 \\
    & 0.8 & 114 & 205 & 242 & 284 \\
    & 0.9 & 116 & 200 & 236 & 280 \\
\midrule
\multirow{5}{2.8cm} {Very low probabilities} & 0.001 & 115 & 195 & 236 & 279 \\
    & 0.005 & 112 & 201 & 243 & 280 \\
    & 0.01 & 115 & 200 & 238 & 279 \\
    & 0.05 & 112 & 201 & 243 & 280 \\
    & 0.1 & 122 & 203 & 249 & 297 \\
  \bottomrule
  \end{tabular}
  \end{center}
\end{table}

We re-run our approach with different factor values.  
The result is shown in \tabref{tab:parameter}. The first column denotes the parameter type. The first five rows represent the five sets of experiments to study the influence of the moderate factor values for
insensitive operations and the last five rows are five sets of experiments to study the very low factor values for sensitive operations. The second column denotes the factor values, where the value represents $P(Result = 1 | Parents = 0)$. 
When studying the effect of changing one type of parameter, we use the default value for the other type. 
We can see that the choice of parameters has only a small impact on the results and our default parameters are not in an optimal position. 
This suggests that our model is robust with respect to different parameters, and could still work without fine-tuning.

\begin{Summary}{}{}
The above results show that the factor values do not have much effect on  SmartFL and the default parameters of SmartFL are not in an optimal position.
\end{Summary}


\subsubsection{RQ5: Combining with other Techniques}
\begin{table}[t]
  \caption{Integrating CombineFL with SmartFL}
  \label{tab:combine}
  \begin{center}
    \begin{tabular}{c|ccc}
    \toprule
    Technique & Top-1 & Top-3 & Top-5 \\
    \midrule
    CombineFL & 71(20\%) & 130(36\%) & 160(45\%) \\
    \midrule
    TRANSFER-FL & 76(21\%) & 133(37\%) & 155(43\%) \\
    \midrule
    \rowcolor{tablegray}
    CombineFL with SmartFL & \textbf{80(22\%)}& \textbf{147(41\%)} & \textbf{168(47\%)} \\
  \bottomrule
  \end{tabular}
  \end{center}
\end{table}
\tabref{tab:combine} shows the results of combining SmartFL with CombineFL. As we can see, combining SmartFL with CombineFL can comprehensively improve the effectiveness of CombineFL as the
results are
better than TRANSFER-FL and the original CombineFL. 

\begin{Summary}{}{}
This experiment shows that SmartFL can be effectively combined with other fault localization methods and improve the effect of the combined method.
\end{Summary}



\subsection{Threats to Validity}
\smalltitle{Internal Validity} The potential threat to internal validity mainly lies in the implementation of SmartFL and the experimental scripts. This may cause inaccurate experimental results or fail to truly reflect the approach of SmartFL. To mitigate the threat, we manually check our code and the logs during experiments. 

\smalltitle{External Validity} The primary threat to external validity is the benchmarks used in our study. To mitigate the threat, we choose the widely-used real-world dataset Defects4J 2.0 as our benchmark. In addition, we conduct experiments on the 835 cases from all 17 projects in Defects4J 2.0 to minimize the threat.

\smalltitle{Construct Validity} The primary threat to construct validity is the metrics we used in our study. To mitigate the threat, we use the widely accepted Top-k metrics, which are commonly utilized in prior fault localization studies. 

\section{Discussion \label{sec:discussion}}
\subsection{The Generalizability of Our Approach}
The implementation and evaluation of our approach are focused solely on Java programs, which may cast doubts on the generalizability of our approach. We acknowledge that in order to implement our approach in other languages, it is necessary to implement corresponding dynamic tracing and building dynamic dependency graphs, which may require a certain amount of engineering overhead.
However, at the approach level, our approach is independent of the Java language because it is a process of converting from a dynamic dependency graph to a Bayesian network. Therefore, in theory, our approach can be extended to any assignment-based imperative programming language. 

\subsection{The Complexity of the Solution}
Since our approach relies on detailed tracing to capture dynamic dependencies and maintain a Bayesian network system, the implementation of our approach may be challenging. This is indeed a shortcoming of our approach, but we believe that although our approach has a certain complexity in implementation, it is stable and valuable after implementation, so it is a worthwhile one-time effort. For example, we have already implemented it robustly and effectively on Java and it has achieved good performance in both effectiveness and efficiency on various programs.  Java developers could use our implementation for various fault localization cases out of the box.

\subsection{The Assumptions in Probabilistic Modeling}
\secref{sec:overview} introduces the core assumption of our paper: ``A faulty evaluation results in an evenly distributed random result. Specifically, an evaluation of an expression is faulty only if one of the three conditions is satisfied'', which may oversimplify real-world fault scenarios. However, there is a trade-off between the complexity of the model and efficiency, and our approach tries to strike a balance. 
There are indeed corner cases where our model is not accurate, but our empirical evaluation on the 835 faults from the 17 Defects4J projects shows that our design is both accurate and efficient in practice.

\subsection{Complexity Analysis}
The time and memory costs of our approach scale linearly with the lengths of the test executions, which indicates that our approach has the potential to scale to even larger programs.

As we discussed in \secref{sec:efficiency}, 
SmartFL consists of three steps: (a) profiling, (b) tracing, and (c) modeling. For profiling, the time overhead is approximately the original test execution time multiplied by a constant as our instrumentation simply records the method coverage.
Similarly, the memory consumption is approximately linear in the test execution time as the memory consumption of information recorded per method is constant.
For tracing, the time overhead is approximately the execution time of the chosen test cases multiplied by a constant since our instrumentation simply records the instructions that are executed.
Similarly, the memory consumption is linear in the number of instructions executed in the chosen test cases as the memory consumption of information recorded per instruction is constant.
For modeling, it consists of building the graph and performing inference on the graph, whose complexities we discuss in detail next.

For building the graph in modeling, the time consumption is dominated by parsing the traces to build the graph while the time consumption of the static analysis is negligible, as our static analysis is lightweight. 
As explained in \secref{sec:implementation}, the parsing process can be viewed as simulating the program execution but only to capture the def-use relationship between runtime values and instructions.
For each instruction, the computation consists of 1) finding nodes in the graph that correspond to the values it accesses, and 2) adding edges between these nodes and the node representing the instruction's correctness based on the dependency introduced by the instruction. 
Both kinds of computation consume constant time so the overall time scales linearly with the sum of the trace lengths.
Similarly, the space consumption scales linearly with the graph size, which is linear in terms of the sum of the trace lengths.

For probabilistic inference in modeling, the probabilistic inference algorithm we apply is an iterative approximate one.
The number of computations in one iteration increases linearly with the size of the graph, and the size of the graph is also linearly related to the length of the considered traces. In addition, we set a maximum number of allowed iterations. Therefore, the time consumption for probabilistic inference will also increase linearly with the length of the test executions.
As for memory consumption, it is linear in the graph size which is in turn linear in the length of the test executions. 

According to the above analysis, the time and space consumption of all steps in our approach increases linearly with the lengths of the test executions, which
indicates that our approach has the potential to scale to even larger programs.

\section{Related Work \label{sec:related}}
Fault localization has been intensively studied during the past decades. Here we leave a full summary of fault localization to respective surveys~\cite{wong2016survey,alipour2012automated,perez2014survey} and empirical studies~\cite{CombineFL}, and discuss only the most related work.

\subsection{Probabilistic Approaches}
Fault localization is inherently a probabilistic analysis process, and many existing approaches resort to probabilistic modeling. Similar to our approach, these modeling approaches also treat the sample space as all possible faults or all possible fault locations and try to identify the element that has the highest conditional probability of being faulty based on the observed test results. Different from our approach which extracts the probabilities from the semantics of the program, most probabilistic approaches 
either consider only the coverage and do not model the semantics of the program~\cite{DBLP:conf/aaai/Gonzalez-SanchezAGG11,DBLP:journals/tse/PerezAD21}.
Some probabilistic approaches learn the probabilities from test executions but do not build precise dynamic dependency graphs ~\cite{DBLP:conf/nips/DietzDZS09,DBLP:journals/tse/BaahPH10}. Kang et al.~\cite{DBLP:conf/issta/KangCY23} proposed a Bayesian framework to explain existing fault localization approaches like SBFL from a probabilistic angle but it does not model the semantics of the program or propose new fault localization approaches.

The only exception is the approach proposed by Xu et al.~\cite{zhangxiangyu}. This approach solves a different problem, namely interactive fault localization: how to support the developer when he faces one failing test. Similar to our approach, their approach also uses probabilistic modeling and introduces Bernoulli probabilistic variables to represent whether the run-time values and statements are faulty or not.
Our approach is inspired by their work, but the fundamental difference is that our approach addresses the scalability challenge while their work cannot. Their work only handles a single failing test of a small program, and cannot be used for non-interactive fault localization. Each scaling technique in our approach is shown to be critical in \secref{sec:evaluation}.

\subsection{Spectrum-based Fault Localization}
As mentioned before, the most popular type of coverage-based fault localization techniques are SBFL approaches, which calculate the suspicious scores of program elements based on the numbers of pass/fail tests covering the element using different formulae~\cite{jones2005empirical,abreu2006evaluation,naish2011model}.


As mentioned, coverage is only one of the four conditions for a test to fail on a fault, and thus the coverage-based approaches do not consider the latter three conditions. To overcome this problem, some existing approaches combine SBFL with program slicing~\cite{reis2019demystifying,mao2014slice,ju2014hsfal}, in the sense that only the statements in the slice can produce and propagate the faults. For example, Mao et al. ~\cite{mao2014slice} proposed SSFL (slice-based statistical fault localization). By calculating the suspiciousness score on an approximate dynamic backward slice, SSFL significantly boosts all 16 formulas of SBFL. However, slicing only reveals the possibility but not the probability that a program element produces or propagates fault, and thus is a very inaccurate modeling of semantics. 
Some other approaches try to make better use of coverage information. Zhang et al.~\cite{DBLP:journals/tosem/ZhangLSYMY23} proposed UNITE to optimize coverage information based on the frequency of each statement appearing in each test and also combine with program slicing. However, these approaches are still limited by the limitations of coverage information and essentially do not accurately model semantics.



\subsection{Approaches Modeling Semantics}
As we have discussed in Sections~\ref{sec:intro} and \ref{sec:overview}, MBFL~\cite{mbfl_moon2014ask,mbfl_papadakis2015metallaxis} and angelic debugging~\cite{chandra2011angelic,DBLP:conf/tacas/ChristakisHMSW19} are the two main families that model semantics for fault localization, but both have scalability issues due to their precise modeling of the semantics. 
Our evaluation also shows that our approach is more than 100 times faster than MBFL and is much more effective.

In addition, there are some new MBFL approaches in recent years. Delta4Ms~\cite{Delta4Ms} integrates the principles of signal theory to model the actual suspiciousness and mutant bias as the desired and false signal components. 
Since Delta4Ms only reports results on 15 cases in the Lang project of Defects4J, and its artifact does not include instructions on how to make it work on the Defects4J cases, we cannot compare with them directly.
LEAM~\cite{leam} leverages DL-based mutation techniques by adapting the syntax-guided encoder-decoder architecture and predicting the statements to be mutated.
Although these approaches have improved a lot compared to traditional MBFL approaches, the scalability issue still remains unsolved, as they still typically need hours to locate a fault.

\subsection{Combination Approaches}
Multiple approaches try to combine existing approaches or different information sources. 
Xuan and Monperrus~\cite{xuan2014learning} proposed a learning-to-rank approach to integrate the suspiciousness scores of 25 existing SBFL formulae. Zou et al.~\cite{CombineFL} further extended this approach to integrate the suspiciousness scores produced by different families of fault localization techniques. Sohn and Yoo~\cite{DBLP:conf/issta/SohnY17} use the learning-to-rank technique to combine the suspiciousness scores of existing SBFL formulae, code complexity metrics, and code history metrics. Li et al.~\cite{DBLP:conf/issta/LiLZZ19} use neural networks to combine suspiciousness scores of SBFL and MBFL, code complexity metrics, and text similarity metrics; K{\"{u}}{\c{c}}{\"{u}}k et al.~\cite{DBLP:conf/icse/KucukHP21} use causal inference techniques and machine learning to integrate predicate outcomes and run-time values. 
Lou et al.~\cite{DBLP:conf/sigsoft/LouZDLSHZZ21} proposed GRACE to leverage graph-based representation learning to embed both the syntax of the program and the coverage information. 
Meng et al.~\cite{DBLP:conf/icse/MengW00022} use BiLSTM-based classifiers to learn deep semantic features of statements and leverage the semantic-based, spectrum-based, and mutation-based features for fault localization by a multi-layer perceptron. CAN ~\cite{DBLP:journals/tse/ZhangLMYXL23} leverage graph neural networks to analyze and combine the failure context for fault localization.
However, none of these approaches are able to model program semantics in detail to infer the probability of introducing and propagating errors during test executions, which is the focus of this paper.

Our approach and the combination approaches are inherently orthogonal, as \secref{sec:evaluation} shows that our approach can be effectively combined into a combination framework~\cite{CombineFL} and significantly improve the
effect. The experiments also show that after combining our method into a relatively old framework, the results can outperform the state-of-the-art combination results.


\subsection{Large Language Models Approaches}
Multiple approaches introduce large language models (LLMs) into fault localization
\cite{DBLP:conf/icse/YangGMH24,DBLP:journals/corr/abs-2308-15276}. These studies show that LLMs have a powerful ability to understand program faults. 
However, the input to these approaches is a relatively small program fragment containing the faulty program element, while other fault localization approaches take the full project source code as input. Therefore, there is still a big difference between the setups in these approaches and practical fault localization scenarios. 
Combining the comprehension of programs from LLMs with our approach is a promising direction, which is future work.

\section{Conclusion \label{sec:conclusion}}
This paper proposes a novel fault-localization method based on the probabilistic graphical model. 
Specifically, we utilize semantic information of different statements, while combining both dynamic and static information into our model. We conduct an experiment on a real-world dataset, Defects4J. Our technique is evaluated to be complementary to existing techniques as it could further improve
state-of-the-art by combining with existing techniques.

To facilitate research, our tool and the fault localization data are available at \repo.


\bibliographystyle{IEEEtran}
\bibliography{references}

\end{document}